\documentclass[a4paper,11pt]{article}
\pdfoutput=1 

\usepackage{jheppub} 
\usepackage[T1]{fontenc} 
\usepackage{subcaption}
\graphicspath{ {graphs/} }

\preprint{YITP-SB-2017-50}

\title{\boldmath \LARGE On Superconformal Four-Point Mellin Amplitudes in Dimension $d>2$}

\author[]{Xinan Zhou}

\affiliation[]{C. N. Yang Institute for Theoretical Physics, \\Stony Brook University, Stony Brook, 11794, NY, USA}

\emailAdd{xinan.zhou@stonybrook.edu}


\abstract{ We present a universal treatment for imposing superconformal constraints on Mellin amplitudes for $\mathrm{SCFT_d}$ with $3\leq d\leq 6$. This leads to a new technique to compute holographic correlators, which is similar but complementary to the ones introduced in \cite{Rastelli:2016nze,Rastelli:2017udc}. We apply this technique to theories in various spacetime dimensions. In addition to reproducing known results, we obtain a simple expression for next-next-to-extremal four-point functions in $AdS_7\times S^4$. We also use this machinery on  $AdS_4\times S^7$ and compute the first holographic one-half BPS four-point function. We extract the anomalous dimension of the R-symmetry singlet double-trace operator with the lowest conformal dimension and find agreement with the 3d $\mathcal{N}=8$ numerical bootstrap bound at large central charge.

 }

\begin{document}
\maketitle
\flushbottom
\section{Introduction}

The computation of $n$-point one-half BPS holographic correlators is in general highly non-trivial. For $n\leq 3$, the form of correlators is fully determined by conformal (and R-symmetry) covariance and such correlators are somewhat less interesting. Starting from $n= 4$, invariant cross ratios  can be formed from the coordinates, and the $n$-point correlator is in general a non-trivial function of such cross ratios. The dependence on these cross ratios encodes rich dynamical information of the bulk theory and cannot be determined just from symmetries alone.  Computationally, the holographic calculation becomes most tractable in the limit when the bulk theory reduces to a weakly coupled supergravity theory. In this limit, there exists a straightforward algorithm based on Witten diagram expansion that allows one to compute in principle any correlation function (see \cite{Rastelli:2017udc} for a review of the traditional approach). However this algorithm is rather cumbersome to carry out in practice and very few explicit examples have been computed in the past two decades following it \cite{DHoker:1999pj,Arutyunov:2000py,Arutyunov:2002ff,Arutyunov:2002fh,Arutyunov:2003ae,Uruchurtu:2007kq,Berdichevsky:2007xd,Uruchurtu:2008kp,Uruchurtu:2011wh}. Four-point functions at tree level are sufficient to exemplify why this method runs out of steam quickly. The main difficulties come from two sources. First, to perform the Witten diagram expansion one needs to obtain all the relevant cubic and quartic vertices by expanding the effective Lagrangian to the quartic order. However this expansion is so complicated that only the case of IIB supergravity on $AdS_5\times S^5$ has been performed in full generality \cite{Arutyunov:1999fb}. Second, the traditional method requires the evaluation of a large number of exchange Witten diagrams whose number generally grows with the Kaluza-Klein levels of the external legs.\footnote{More precisely, the number of exchange Witten diagrams grows with the {\it extremality} of the four-point function which is proportional to $\Delta_2+\Delta_3+\Delta_4-\Delta_1$. Here we assumed that the conformal dimension $\Delta_1$ is the largest. See Section \ref{6dapplication} for a detailed discussion.} These Witten diagrams involve non-trivial integrals which are hard to evaluate in position space. For certain spectrums (such as for $AdS_5\times S^5$ and $AdS_7\times S^4$) all the requisite exchange Witten diagrams can be written as a finite sum of contact diagrams using the method of \cite{DHoker:1999aa}. For other models such as eleven dimensional supergravity compactified on $AdS_7\times S^4$, however, this trick does not apply and even the simplest exchange Witten diagrams cannot be expressed as such a finite sum.

Recently, modern bootstrap-inspired methods have been introduced to make the computation of holographic correlators more efficient. These methods point out that the organizational principles of holographic correlators are symmetries and self-consistency conditions, and the precise knowledge of the full effective Lagrangian is unnecessary. In \cite{Rastelli:2016nze,Rastelli:2017udc} two such methods were presented. The first method was dubbed the position space method and mimics the traditional algorithm of computing correlators. To implement this method, one writes down an ansatz for the correlator as a linear combination of all the Witten diagrams. The crucial departure from the traditional method is that instead of obtaining these coefficients using bulk vertices of the effective Lagrangian, one leaves them as unfixed parameters. One then imposes the superconformal Ward identity on the ansatz, which gives enough equations to fix all these parameters. A technical requirement of this approach is that all the exchange Witten diagram should be expressible using a finite number of contact Witten diagrams. This is indeed true for $AdS_5\times S^5$ and $AdS_7\times S^4$ but no longer holds for $AdS_4\times S^7$. Moreover this method also contains a step to reduce all the contact diagrams into a basis. The decomposition becomes increasingly time consuming as the dimensions of the external operators grow and therefore is not suitable for such heavy four-point functions.

The second method we introduced is more elegant and takes advantage of the Mellin amplitude technology initiated by Mack \cite{Mack:2009mi}. Computational difficulties aside, the position space calculation yields final results as complicated sums of contact diagrams which look totally obscure. When translated into Mellin space, however, the answers simplify significantly and the scattering nature of the correlators becomes manifest. This suggests that Mellin space is the correct language for holographic correlators and one should try to avoid the excursion into position space altogether. Moreover, as we have learned from the modern on-shell methods of scattering amplitude in flat space (see, {\it e.g.}, \cite{Elvang:2015rqa,nima} for recent textbook presentations), it is not the best strategy to compute the full amplitude by summing up all the Feynman diagrams. Instead, by exploiting consistency conditions such as dimensional analysis, Lorentz invariance and locality, the correct S-matrix  can be sculpted out directly.  Since  holographic correlators are essentially just on-shell supergravity amplitudes in a maximally symmetric bulk spacetime, it is our prejudice that a similar story should exist for correlation functions and we can obtain the full correlator without resorting to individual Witten diagrams. In our second approach, we achieved precisely that. We translate the task of computing holographic four-point functions into solving an algebraic bootstrap problem in Mellin space. The keystone of this method is to rephrase the position space solution to the superconformal Ward identity in Mellin space. It eventually boils down to a difference operator which repackages the Mellin amplitude into a simpler auxiliary amplitude. This structure of the Mellin amplitude is then combined with other consistency conditions (analytic structures, Bose symmetry and flat-space limit) to form a highly restrictive bootstrap problem. The problem for IIB supergravity compactified on $AdS_5\times S^5$ has been solved and reveals a remarkably simple answer \cite{Rastelli:2016nze,Rastelli:2017udc}. A similar analysis can be performed to set up a bootstrap problem for eleven dimensional supergravity on $AdS_7\times S^4$ \cite{longads7}. Unfortunately, this method  also seems to fall short of the case of $AdS_4\times S^7$, because the position space solution to the 3d $\mathcal{N}=8$ superconformal Ward identity contains nonlocal differential operators. Such operators do not have a clear interpretation in Mellin space.

In this paper we propose another approach which combines elements drawn from the above two methods with new techniques that we will introduce here. This new way of computing holographic correlators inherits the advantage of the Mellin space method of working exclusively inside the Mellin space, but is not subject to the constraint of special spacetime dimensions. Similar to the position space method, this approach is also based on making a suitable ansatz (for the Mellin amplitude) and then solving the ansatz using  superconformal symmetry. Our key ingredient is to have a universal treatment for the superconformal constraints on Mellin amplitudes for SCFTs in all dimensions greater than two.\footnote{According to a classic result of Nahm \cite{Nahm:1977tg}, superconformal field theories can only exist in spacetime dimension less than or equal to six. Therefore our techniques applies to theories with $3\leq d\leq 6$. } For $d>2$, the superconformal Ward identity for one-half BPS four-point functions takes the universal form \cite{Dolan:2004mu}
\begin{equation}
(\chi\partial_\chi-\epsilon\alpha\partial_\alpha)\mathcal{G}(\chi,\chi';\alpha,\alpha')\big|_{\alpha=1/\chi}=0\;
\end{equation}
where $\epsilon=d/2-1$. The variables $\chi$, $\chi'$, $\alpha$, $\alpha'$ are related to the usual conformal and R-symmetry cross ratios by (\ref{changeofvariable}) and $\mathcal{G}(\chi,\chi';\alpha,\alpha')$ is the four-point correlator with certain powers of the position and R-symmetry coordinates stripped off (see Section \ref{scfwi} for details). We will show how this superconformal Ward identity in position space can be used to derive constraints on the Mellin amplitudes -- they are the Mellin space superconformal Ward identities. We will discuss in detail two applications. One is the stress-tensor multiplet four-point function for $AdS_4\times S^7$. With a Mellin amplitude ansatz similar to the position space ansatz, we show that the amplitude is uniquely fixed, up to rescaling, after we impose crossing symmetry and the Mellin space superconformal Ward identities. The last constant is determined by further using a supergravity three-point function. It is worth pointing out that this correlator -- the simplest one-half BPS four-point function for $AdS_4\times S^7$ -- has never been computed in the literature and is accessible to neither method from \cite{Rastelli:2016nze,Rastelli:2017udc}. We also apply this method to study next-next-to-extremal four-point functions in $AdS_7\times S^4$. Although these correlators are in principle subject to our position space method, only a small sector with small external conformal dimensions can be probed in practice because of computational difficulties. This problem is not present in our new method in Mellin space.  In solving such four-point functions, we use the {\it most general} ansatz for the Mellin amplitude that is compatible with the qualitative features of the bulk supergravity, namely, the spectrum, cubic coupling selection rules and the flat space limit. In particular, {\it no} Witten diagram needs to be computed. The evaluation of these diagrams would otherwise be rather lengthy in position space. We show that the superconformal symmetry uniquely determines all the ratios of the coefficients. We also reveal a remarkable simplicity in the final answer: the full Mellin amplitude of all next-next-to-extremal correlators can be repackaged into only one term using a difference operator. While we focus our attention on these specific examples, we expect the method and philosophy in our paper to apply more widely.

This paper is divided into two parts. In part I, we explain how to obtain superconformal constraints on the Mellin amplitudes from the position space superconformal Ward identity. Using this technique we propose a new way to compute holographic four-point functions that works for all dimensions.  In part II, we apply this method to study next-next-to-extremal four-point functions in $AdS_7\times S^4$ and the four-point function of the stress-tensor multiplet in $AdS_4\times S^7$. We also extract the anomalous dimension of the lowest-dimension, R-symmetry singlet, double-trace operator from the $AdS_4\times S^7$ four-point correlator and compare it with the numeric 3d $\mathcal{N}=8$ conformal bootstrap bound. We conclude the paper with a brief discussion.

\section{Part I: Superconformal Ward Identity for Mellin Amplitudes}\label{partI}
 In the first part of this paper we give a universal treatment for imposing superconformal constraints on Mellin amplitudes that works for all dimensions $3\leq d\leq 6$. We start with  reviewing some preliminaries. In Section \ref{scfwi} we review the superconformal Ward identity for one-half BPS four-point functions in the position space representation, and we introduce the Mellin representation in Section \ref{mellinrep}. In Section \ref{solscfwi} we translate the position space superconformal Ward identity into identities in the Mellin space. We then describe in Section \ref{bootstrapmellin} an algorithm to compute four-point functions in the supergravity limit and use the stress-tensor multiplet four-point function for $AdS_5\times S^5$ as a simple demonstration. 

\subsection{Superconformal Ward Identity  in Position Space for $\mathrm{SCFT_{d>2}}$}\label{scfwi}
In this paper we focus on the four-point functions of the one-half BPS operators. These operators are the superprimaries of the shortened representations of superconformal algebras in $d=3,4,5,6$ dimensions, annihilated precisely by half of the supercharges. We will further restrict ourselves to the subset of superconformal algebras for which the R-symmetry group is locally isomorphic to an $SO(n)$ group.\footnote{They include 3d $OSp(4|\mathcal{N})$ ($\mathcal{N}$ even), 4d $(P)SU(2,2|\mathcal{N})$ with $\mathcal{N}=2,4$, 5d $F(4)$ and 6d $OSp(8^*|2\mathcal{N})$ with $\mathcal{N}=1,2$.} For such algebras, the one-half BPS operators $\mathcal{O}_k^{I_1\ldots I_k}$ are in the rank-$k$ symmetric-traceless representation of $SO(n)$ and have quantized conformal dimension
\begin{equation}
\Delta=\epsilon k\;,\quad\quad\quad \epsilon\equiv \frac{d}{2}-1\;.
\end{equation}
A  familiar example is the $d=4$ $\mathcal{N}=4$ SYM theory which has $n=6$. The one-half BPS operators $\mathcal{O}_k^{I_1\ldots I_k}$ are the  single-trace operators ${\rm tr} X^{\{I_1}\ldots X^{I_k\}}$ made out of the six scalars $X^I$, and the operator $\mathcal{O}_k^{I_1\ldots I_k}$ has protected conformal dimension $k$. 

It is convenient to take care of the R-symmetry indices by contracting them with auxiliary null vectors $t^I$
\begin{equation}
\mathcal{O}_k(x,t)\equiv \mathcal{O}_k^{I_1\ldots I_k}(x)\,t_{I_1}\ldots t_{I_k}\;,\quad\quad t^I t_I=0\;.
\end{equation}
Then the four-point function of such one-half BPS operators is index-free and depends on both the spacetime coordinates $x_i$ and the internal R-symmetry coordinates $t_i$
\begin{equation}
G(x_i,t_i)\equiv \langle\mathcal{O}_{k_1}(x_1,t_1)\mathcal{O}_{k_2}(x_2,t_2)\mathcal{O}_{k_3}(x_3,t_3)\mathcal{O}_{k_4}(x_4,t_4)\rangle\;.
\end{equation}
 Define $t_{ij} \equiv t_i\cdot t_j$, the R-symmetry covariance and null property require that the $t_i$ variables can only appear as sum of monomials $\prod_{i<j} (t_{ij})^{\gamma_{ij}}$ where the powers $\gamma_{ij}$ are non-negative integers. Moreover, in order to have the correct scaling behavior when independently rescaling each null vector $t_i\to \zeta_i t_i$, the exponents $\gamma_{ij}$ need to be further constrained by the condition $\sum_{i\neq j}\gamma_{ij}=k_j$. This set of constraints is solved  with the following parameterization,
\begin{equation} \label{para}
\begin{split}
\gamma_{12}={}&-\frac{a}{2}+\frac{k_1+k_2}{2}\;,\;\;\;\;\;\;\;\;\;\;\;\;\gamma_{34}=-\frac{a}{2}+\frac{k_3+k_4}{2}\;,\\
\gamma_{23}={}&-\frac{b}{2}+\frac{k_2+k_3}{2}\;,\;\;\;\;\;\;\;\;\;\;\;\;\gamma_{14}=-\frac{b}{2}+\frac{k_1+k_4}{2}\;,\\
\gamma_{13}={}&-\frac{c}{2}+\frac{k_1+k_3}{2}\;,\;\;\;\;\;\;\;\;\;\;\;\;\gamma_{24}=-\frac{c}{2}+\frac{k_2+k_4}{2}\;,\\
\end{split}
\end{equation}
with the additional condition $a+b+c=k_1+k_2+k_3+k_4$.

\begin{figure}[htbp]
\begin{center}
\includegraphics[scale=0.45]{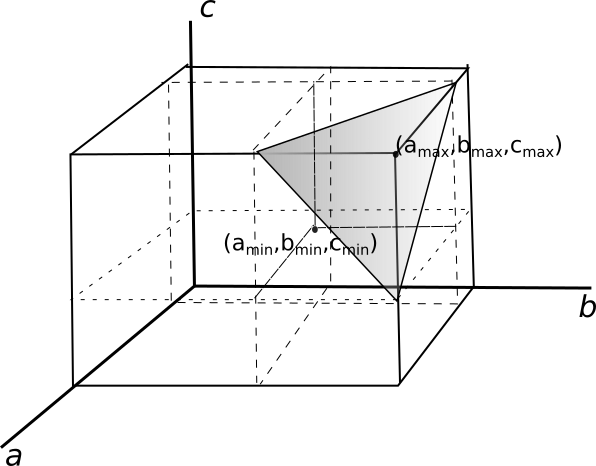} 
\caption{Solution to the $\gamma_{ij}$ constraints.}
\label{cube}
\end{center}
\end{figure}

We can assume $k_1\geqslant k_2\geqslant k_3\geqslant k_4$ without loss of generality. It leaves us with two possibilities, namely,
\begin{equation}
k_1+k_4\leqslant k_2+k_3  \quad ({\rm case \; I}) \quad {\rm and} \quad
 k_1+k_4> k_2+k_3 \quad ({\rm case \; II}) \,.
\end{equation}
The inequality constraints $\gamma_{ij} \geqslant 0$  
define in either case a cube inside the parameter space  $(a,b,c)$, as shown in Figure \ref{cube}. The condition $a+b+c=k_1+k_2+k_3+k_4$ further restricts the solution to be the equilateral triangle inside the cube shown shaded in the figure. We denote the coordinates of  vertices of the cube closest and furthest from the origin as $(a_{\mathrm{min}},b_{\mathrm{min}},c_{\mathrm{min}})$ and $(a_{\mathrm{max}},b_{\mathrm{max}},c_{\mathrm{max}})$.  Then 
\begin{equation}
\begin{split}
a_{\mathrm{max}}=k_3+k_4\;,\;\;\; {}& a_{\mathrm{min}}=k_3-k_4\;, \quad\quad\quad\quad\quad a_{\mathrm{max}}=k_3+k_4\;,\;\;\; {} a_{\mathrm{min}}=k_1-k_2\;,\\
b_{\mathrm{max}}=k_1+k_4\;,\;\;\; {}& a_{\mathrm{min}}=k_1-k_4\;, \quad\quad\quad\quad\quad b_{\mathrm{max}}=k_2+k_3\;,\;\;\; {} a_{\mathrm{min}}=k_1-k_4\;,\\
c_{\mathrm{max}}=k_2+k_4\;,\;\;\; {}& a_{\mathrm{min}}=k_2-k_4\;,\quad\quad\quad\quad\quad c_{\mathrm{max}}=k_2+k_4\;,\;\;\; {} a_{\mathrm{min}}=k_1-k_3\;.\\
\text{(case I)}& \quad\quad\quad\quad\quad\quad\quad\quad\quad\quad\quad\quad\quad\quad\quad\quad\quad\quad  \text{(case II)}
\end{split}
\end{equation}
Let $2\mathcal{L}$ be the length of each side of the cube, we find
\begin{equation}
{\mathcal{L}}={} k_4\quad \text{(case I)}\;,\quad\quad\quad\quad
{\mathcal{L}}={} \frac{k_2+k_3+k_4-k_1}{2}\quad \text{(case II)}\;.
\end{equation} 
It is clear from the parametrization  (\ref{para})  that $\gamma_{ij}$ has lower bounds  $\gamma_{ij} \geq \gamma_{ij}^0$. These $\gamma_{ij}^0$ are obtained by
substituting the maximal values $(a_{\mathrm{max}},b_{\mathrm{max}},c_{\mathrm{max}})$,
\begin{equation}
\begin{split}
\gamma^0_{12}={}& \frac{k_1+k_2-k_3-k_4}{2}\;,\\
\gamma^0_{13}={}& \frac{k_1+k_3-k_2-k_4}{2}\;,\\
\gamma^0_{34}={}&\gamma^0_{24}=0\;,\\
\gamma^0_{14}={}&0\;\;\;\;\mathrm{(case\;\; I)},\;\;\;\;\;\;  \frac{k_1+k_4-k_2-k_3}{2}\;\;\mathrm{(case\;\; II)}\;,\\
\gamma^0_{23}={}& \frac{k_2+k_3-k_1-k_4}{2}\;\;\;\;\mathrm{(case\;\; I)},\;\;\;\;\;\; 0\;\;\mathrm{(case\;\; II)}\;.
\end{split}
\end{equation}
We now factor out the product $\prod_{i<j}\left(\frac{t_{ij}}{x_{ij}^{2\epsilon}}\right)^{\gamma^0_{ij}}$ from the correlator -- each $\left(\frac{t_{ij}}{x_{ij}^{2\epsilon}}\right)^k$ is the two-point function of a weight-$k$ one-half BPS operator. The object we obtain has the  scaling behavior of a
four-point function with  equal weights $\mathcal{L}$. This behavior further motivates us to define
\begin{equation}\label{GandcurlyG}
G(x_i,t_i)=\prod_{i<j}\left(\frac{t_{ij}}{x_{ij}^{2\epsilon}}\right)^{\gamma^0_{ij}}\left(\frac{t_{12}t_{34}}{x^{2\epsilon}_{12}x^{2\epsilon}_{34}}\right)^\mathcal{L} \mathcal{G}(U,V;\sigma,\tau)\;.
\end{equation}
The outstanding factors take care of the covariance under the conformal and R-symmetry group, and the correlator is reduced into a function $\mathcal{G}(U,V;\sigma,\tau)$ depending on only four invariant variables. Here we used the usual conformal cross ratios
\begin{equation}
      U = \frac{(x_{12})^2(x_{34})^2}{(x_{13})^2(x_{24})^2}\;,\;\;\;\;\;\;\;\; V =   \frac{(x_{14})^2(x_{23})^2}{(x_{13})^2(x_{24})^2}
\end{equation}
and analogously the R-symmetry cross ratios 
\begin{equation}
       \sigma = \frac{(t_{13}) (t_{24})}{(t_{12}) (t_{34})}\;,\;\;\;\;\;\;\;\;\;\;\;\;\;\;\;\tau=  \frac{(t_{14}) (t_{23})}{(t_{12}) (t_{34})}\;.
\end{equation}
It is not difficult to see that $\mathcal{G}(U,V;\sigma,\tau)$ is a polynomial of $\sigma$ and $\tau$ with degree $\mathcal{L}$. 

So far we have only required the correlator to be covariant under the bosonic part of the superconformal group. The fermionic generators impose further constraints on $\mathcal{G}(U,V;\sigma,\tau)$ in the form of a {\it superconformal Ward identity}. It is useful to make a change of variables
\begin{equation}
\begin{split}
{}&U=\chi \chi'\;,\quad\quad\quad\quad\quad\quad V=(1-\chi)(1-\chi')\;,\\
{}&\sigma=\alpha \alpha'\;,\quad\quad\quad\quad\quad\quad \tau=(1-\alpha)(1-\alpha')\;.
\end{split} \label{changeofvariable}
\end{equation}
In terms of these variables, the superconformal Ward identity takes the universal form \cite{Dolan:2004mu}
\begin{equation}\label{scfwiposition}
(\chi\partial_\chi-\epsilon\alpha\partial_\alpha)\mathcal{G}(\chi,\chi';\alpha,\alpha')\big|_{\alpha=1/\chi}=0\;.
\end{equation}

\subsection{Mellin Representation}\label{mellinrep}
We now review some essential features of the Mellin amplitude formalism introduced by Mack \cite{Mack:2009mi} and developed in \cite{Penedones:2010ue,Paulos:2011ie,Fitzpatrick:2011ia,Costa:2012cb,Fitzpatrick:2012cg,Costa:2014kfa,Goncalves:2014rfa}.\footnote{For other applications and recent developments, see \cite{Paulos:2012nu,Nandan:2013ip,Lowe:2016ucg,Rastelli:2016nze,Paulos:2016fap,Nizami:2016jgt,Gopakumar:2016cpb,Gopakumar:2016wkt,Dey:2016mcs,Aharony:2016dwx, Rastelli:2017ecj,Dey:2017fab, Yuan:2017vgp,Dey:2017oim,Faller:2017hyt,Chen:2017xdz}). } We begin with a general $n$-point correlation function of scalar operators with conformal dimensions $\Delta_i$. Using conformal symmetry, we can write the correlator as
\begin{equation}
G_{\Delta_1,\ldots,\Delta_n}(x_1,\ldots,x_n)=\prod_{i<j}(x^2_{ij})^{-\delta_{ij}^0}\mathcal{G}(\xi_r) \, ,
\end{equation}
where $\xi_r$ are the conformally invariant cross ratios of the form $x_{ij}^2x_{kl}^2x_{il}^{-2}x_{kj}^{-2}$. For the right hand side to transform with appropriate weights under conformal transformations, the powers $\delta_{ij}^0$ must satisfy the constraints
\begin{equation} \label{deltaconstraints}
\sum_{j\neq i}\delta_{ij}^0=\Delta_i\;.
\end{equation}
These constraints admit $\frac{1}{2}n(n-3)$ solutions, in correspondence with the $\frac{1}{2}n(n-3)$ cross ratios (here we ignore the algebraic relations that arises for small $d$).\footnote{More generally, the number of independent cross ratios  is $nd-\frac{1}{2}(d+1)(d+2)$ if $n\geqslant d+1$ and $\frac{1}{2}n(n-3)$ if $n< d+1$. See, {\it e.g.}, \cite{Rastelli:2017udc} for details of the counting.} 

Instead of taking $\delta^0_{ij}$ to be fixed,  Mack \cite{Mack:2009mi} suggested we should view them as variables $\delta_{ij}$ obeying the same constraints, 
\begin{equation}\label{deltaconstr}
\delta_{ij} = \delta_{ji}  \, ,\quad \sum_{j}\delta_{ij}=\Delta_i \, , 
\end{equation}
and write 
the correlator as an integral transform with respect to these variables. More precisely, one defines the following (inverse) Mellin transform for the {\it connected}\footnote{The disconnected part is a sum of powers of $x_{ij}^2$ and
its Mellin transform is ill-defined.} part of the correlator,
\begin{equation}\label{mack}
G_{\Delta_1,\ldots,\Delta_n}^{\rm conn}(x_1,\ldots,x_n)=\int [d\delta_{ij}] M(\delta_{ij})\prod_{i<j}(x^2_{ij})^{-\delta_{ij}}
\end{equation}
The integration is performed with respect to the $\frac{1}{2}n(n-3)$ independent variables along the imaginary axis. The correlator $\mathcal{G}(\xi_r)_{\rm conn}$ is captured by the function $M(\delta_{ij})$ which is the {\it reduced} Mellin amplitude following Mack's terminology.

We can solve constraints (\ref{deltaconstr}) by introducing some auxiliary  ``momentum'' variables $p_i$ which live in a $d+1$-dimensional spacetime,
\begin{equation}
\delta_{ij}=p_i\cdot p_j\;.
\end{equation}
These momentum variables obey ``momentum conservation'' and the ``on-shell'' condition
\begin{equation}
\sum_{i=1}^n p_i=0\;,\quad\quad\quad\quad\quad p_i^2=-\Delta_i\;.
\end{equation}
The number of  independent  ``Mandelstam variables'' $\delta_{ij}$  is then also $\frac{1}{2}n(n-3)$.\footnote{A more careful counting (see, {\it e.g.}, \cite{Rastelli:2017udc}) reveals that in a $D$-dimensional spacetime the number of such independent variables is $n(D-1)-\frac{1}{2}D(D+1)$ if $n\geqslant D$ and $\frac{1}{2}n(n-3)$ if $n<D$.  We conclude that the counting of independent Mandelstam variables in $D$ dimensions coincides precisely with the counting of independent conformal cross ratios in $d$ dimensions if we set $D=d+1$.}

The Mellin representation encodes the operator product expansion into simple analytic properties of $M(\delta_{ij})$. 
Indeed, consider the OPE
\begin{equation}
{\cal O}_i (x_i) {\cal O}_j (x_j) = \sum_k  c_{ij}^{\; k} \, \left(  (x_{ij}^2)^{-\frac{\Delta_i + \Delta_j - \Delta_k}{2} } {\cal O}_k (x_k)    \,  + {\rm  descendants} \right) \, ,
\end{equation}
and take ${\cal O}_k$ to be a scalar operator for simplicity. The reduced Mellin amplitude  $M$ must have a pole at $\delta_{ij} =  \frac{\Delta_i + \Delta_j  - \Delta_k}{2}$ in order to reproduce the leading behavior as $x_{ij}^2 \to 0$. This can be easily seen by closing the $\delta_{ij}$ integration contour to the left on the complex plane. More generally,
the location of the leading pole is determined by the twist $\tau$ of the exchanged operator ($\tau \equiv \Delta- \ell$). Conformal descendants contribute an infinite sequence of satellite poles. 
The upshot is that for any primary operator ${\cal O}_k$ of twist $\tau_k$  appearing in the ${\cal O}_i {\cal O}_j$ OPE,
the reduced Mellin amplitude $M(\delta_{ij})$ has poles at
\begin{equation}  \label{Mellinpoles}
\delta_{ij} =  \frac{\Delta_i + \Delta_j  - \tau_k - 2n}{2} \, , \quad n = 0, 1, 2 \dots \, .
\end{equation}
Mack further defined \textit{Mellin amplitude} $\mathcal{M}(\delta_{ij})$ by separating out a product of Gamma functions,
\begin{equation}\label{Mackmellin}
\mathcal{M}(\delta_{ij}) \equiv \frac{M(\delta_{ij})}{\prod_{i<j}\Gamma[\delta_{ij}]}\, .
\end{equation} 
This definition has two advantages. For one, in the large $N$ theories, the poles of the Gamma functions captures the double-trace operators to the $O(1/N^2)$ order in the OPE. Factoring out these Gamma functions cleanly separates the dynamical information of the single-trace operators from the ubiquitous double-trace operators\footnote{More precisely, the explicit Gamma functions provides the exactly needed analytic structure which is anticipated in the large $N$ CFT. As is clear from the expression (\ref{mellinexchnage}) below, the Mellin amplitude $\mathcal{M}$ contains only poles associated with the exchanged single-trace operator. Fixing the single-trace contribution is enough to determine the double-trace contribution. This is most evident in Mellin space, but can also be argued more abstractly from CFT reasonings using crossing symmetry \cite{Caron-Huot:2017vep,Alday:2017gde,Kulaxizi:2017ixa,Li:2017lmh}.}. For the other, especially for four-point functions, the s-channel OPE $(x_{12} \to 0)$ implies
that the Mellin amplitude ${\cal M} (s, t)$ has poles in $s$ with residues that are {\it polynomials} of $t$. These residues are called {\it Mack polynomials} and 
depend on the spin of the exchanged operator. They are in close analogy with the familiar partial wave expansion of  a flat-space S-matrix.

Now we focus on the Mellin amplitudes of four-point functions. The representation (\ref{mack}) can be written more explicitly as
\begin{equation}\label{inverseG}
 \mathcal{G}_{\rm conn}(U,V;\sigma,\tau) = \int_{-i\infty}^{i\infty} \frac{ds}{4\pi i} \frac{dt}{4\pi i}\; U^{\frac{s}{2}-\frac{\epsilon(k_3+k_4)}{2}+\epsilon\mathcal{L}}\,V^{\frac{t}{2}-\frac{\epsilon \min\{k_1+k_4,k_2+k_3\}}{2}}\mathcal{M}(s,t;\sigma,\tau) \Gamma_{k_1k_2k_3k_4}
\end{equation}
where
\begin{equation}\label{gammafactor}
\begin{split}
\Gamma_{k_1k_2k_3k_4} \equiv {}&\Gamma[-\frac{s}{2}+\frac{\epsilon(k_1+k_2)}{2}]\Gamma[-\frac{s}{2}+\frac{\epsilon(k_3+k_4)}{2}]\Gamma[-\frac{t}{2}+\frac{\epsilon(k_2+k_3)}{2}]\\
\times{}&\Gamma[-\frac{t}{2}+\frac{\epsilon(k_1+k_4)}{2}]\Gamma[-\frac{u}{2}+\frac{\epsilon(k_1+k_3)}{2}]\Gamma[-\frac{u}{2}+\frac{\epsilon(k_2+k_4)}{2}]\; , \\
u \equiv &\; \epsilon(k_1+k_2+k_3+k_4)-s-t\,.
\end{split}
\end{equation}
We have substituted in $\Delta_i=\epsilon k_i$ for the conformal dimensions and also restored the R-symmetry variables $\sigma$, $\tau$ for later convenience\footnote{We can also decompose both sides into R-symmetry monomials
\begin{equation}\label{inverseGmono}
 \mathcal{G}_{{\rm conn},LMN}(U,V) = \int_{-i\infty}^{i\infty} \frac{ds}{4\pi i} \frac{dt}{4\pi i}\; U^{\frac{s}{2}-\frac{\epsilon(k_3+k_4)}{2}+\epsilon\mathcal{L}}\,V^{\frac{t}{2}-\frac{\epsilon \min\{k_1+k_4,k_2+k_3\}}{2}}\mathcal{M}_{LMN}(s,t) \Gamma_{k_1k_2k_3k_4}\;,
\end{equation}
and it is related to (\ref{inverseG}) by $\mathcal{G}_{\rm conn}(U,V;\sigma,\tau)=\sum_{L+M+N=\mathcal{L}}\sigma^M\tau^N  \mathcal{G}_{{\rm conn},LMN}(U,V)$ and $\mathcal{M}(s,t;\sigma,\tau)=\sum_{L+M+N=\mathcal{L}}\sigma^M\tau^N  \mathcal{M}_{LMN}(s,t)$.}. The effectiveness of this formalism is best illustrated by its application to the Witten diagrams. To start with, the Mellin amplitude of a four-point contact Witten diagram (known as the $D$-function) is just a constant,
\begin{equation}
D_{\Delta_1\Delta_2\Delta_3\Delta_4}=\int [d\delta_{ij}] \left(\frac{\pi^{d/2}\Gamma[\frac{\sum\Delta_i}{2}-d/2]}{\prod \Gamma[\Delta_i]}\right)\times \prod_{i<j}\Gamma[\delta_{ij}](x^2_{ij})^{-\delta_{ij}}\;.
\end{equation}
Exchange diagrams are also much simpler in Mellin space. The s-channel exchange Witten diagram  with an exchanged field of conformal dimension $\Delta$ and spin $J$  has a Mellin amplitude with the following simple analytic structure \cite{Costa:2012cb},
\begin{equation}\label{mellinexchnage}
\mathcal{M}(s,t)= \sum_{m=0}^{\infty}\frac{Q_{J,m}(t)}{s-\tau-2m}+P_{J-1}(s,t) \, ,
\end{equation}
where $\tau=\Delta-J$ is the twist. Here $Q_{J,m}(t)$ are polynomials in $t$ of degree $J$ and $P_{J-1}(s,t)$ polynomials in $s$ and $t$ of degree $J-1$. These polynomials depend on the dimensions $\Delta_{1,2,3,4}$, $\Delta$, as well as the spin $J$. The singular part of the Mellin amplitude in fact coincides with that of the Mellin amplitude of a conformal block with the same $\Delta$ and $\ell$.

It has been observed (see, {\it e.g.}, \cite{Penedones:2010ue}) that the infinite series of poles in (\ref{mellinexchnage}) truncates to a finite sum if a condition of quantum numbers is met
\begin{equation}\label{trunccond}
 \tau = \Delta_1+\Delta_2 \;\; {\text{mod}}\;\; 2 \;,\quad{\text{or}}\quad    \tau = \Delta_3+\Delta_4\;\; {\text{mod}}\;\; 2,
 \end{equation}
  {\it i.e.}, when the series of simple poles in $\mathcal{M}(s,t)$ start to overlap with the $s$-poles in the Gamma factors. One finds that the maximal value $m_{\rm max}$ of $m$  is given by $ \Delta_1 + \Delta_2 -\tau = 2 (m_{\rm max}+1)$ in the first case
 and by $\Delta_3 + \Delta_4-\tau = 2 (m_{\rm max}+1)$ in the second case. This is the Mellin space  version of a phenomenon first discovered in \cite{DHoker:1999aa}: an exchange Witten diagram with these special values of quantum numbers can be written as a {\it finite} sum of contact Witten diagrams. This remarkable simplification is dictated by the compatibility with the large $N$ OPE in the dual CFT, as was explained in detail in \cite{longads7}.

Last but not the least, one can argue on general grounds that in the large $s$ and $t$ limit, the Mellin amplitude should reduce to the flat-space bulk S-matrix. A precise prescription for relating the massless\footnote{For massive external particles, see the discussion in \cite{Paulos:2016fap}.} flat-space scattering amplitude $\mathcal{T}(P_i)$    to the asymptotic form of the holographic Mellin amplitude   was given in \cite{Penedones:2010ue} and justified in \cite{Fitzpatrick:2011hu},
\begin{equation} \label{joao}
{\cal M} (\delta_{ij}) \approx \frac{R^{n(1-d)/2+d+1}}{\Gamma(\frac{1}{2}\sum_i\Delta_i-\frac{d}{2})}\int_0^\infty d\beta \beta^{\frac{1}{2}\sum_i\Delta_i-\frac{d}{2}-1}e^{-\beta}\mathcal{T}\left(S_{ij}=\frac{2\beta}{R^2}s_{ij}\right)\;. 
\end{equation}
where $S_{ij} = -(P_i + P_j)^2$ are the Mandelstam invariants of the flat-space scattering process. In flat space, we have a precise opinion of the asymptotic behavior of the four-point amplitude ${\cal T}(S, T)$ --  it can grow at most linearly for large $S$ and $T$. Indeed, a spin $\ell$ exchange diagrams grows with power $\ell -1$, and the graviton with $\ell = 2$ is the highest spin state in the single-trace field spectrum. Similarly, contact interactions with $2n$ derivatives give rise to a contribution with power $n$ growth. Since supergravity (in ten-dimensional or eleven dimensional flat space) contains contact interactions with at most two derivatives, the growth of the contact interaction contribution should also be linear.  All in all, we deduce from (\ref{joao}) 
\begin{equation} \label{largelambda}
{\cal M}(\beta s, \beta t) \sim O(\beta) \quad {\rm for} \; \beta \to \infty \,.
\end{equation}

\subsection{Superconformal Ward Identity in Mellin Space}\label{solscfwi}

From the Mellin space point of view the superconformal Ward identity (\ref{scfwiposition})  seems rather unappealing at first sight. Our ideal scenario is to have factors in the form of $U^mV^n$ multiplying an inverse Mellin transformation.  We can absorb such factors by shifting the $s$, $t$ variables and trade them for difference operators that act on the integrand. However, the variables $\chi$ and $\chi'$ appear asymmetrically on the left side of the identity (\ref{scfwiposition}). If one naively solved $\chi$ and $\chi$ in terms of $U$, $V$, one would encounter square roots in these variables, making how to proceed unclear. 

We now offer in this subsection a simple observation. This observation allows us to obtain relations in the Mellin amplitude from the position space identity (\ref{scfwiposition}), and these relations constitute the superconformal Ward identities in Mellin space.  For starters, let us write the differential operator $\chi\partial_\chi$ as
\begin{equation}\label{chidchi}
\chi\frac{\partial}{\partial \chi}=U\frac{\partial}{\partial U}+V\frac{\partial}{\partial V}-\frac{1}{1-\chi}V\frac{\partial}{\partial V}\;.
\end{equation}
We act this operator on $\mathcal{G}_{\rm conn}(U,V;\sigma,\tau)=\sum_{L+M+N=\mathcal{L}}\sigma^M\tau^N  \mathcal{G}_{{\rm conn},LMN}(U,V)$ but do not evaluate the action of $U\partial_U$ and $V\partial_V$ on $\mathcal{G}_{{\rm conn},LMN}(U,V)$ at this stage. For the R-symmetry part, acting with $\alpha\partial_\alpha$ and then setting $\alpha=1/\chi$ turn the monomials $\sigma^i\tau^j$ into some simple rational functions
\begin{eqnarray}
\nonumber1&\to& 0\;,\\
\nonumber\sigma &\to& \left(\frac{1}{\chi}\right)\alpha'\;,\\
\nonumber\tau &\to&  \left(\frac{1}{\chi}\right)\alpha' -\frac{1}{\chi}\;,\\
\nonumber\sigma^2 &\to& \left(\frac{2}{\chi^2}\right)\alpha'^2 \;,\\
\nonumber\sigma\tau &\to&  \left(\frac{2-\chi}{\chi^2}\right)\alpha'^2- \nonumber\left(\frac{2-\chi}{\chi^2}\right)\alpha'\;,\\
\tau^2 &\to&  \left(\frac{2(1-\chi)}{\chi^2}\right)\alpha'^2+ \left(\frac{4(1-\chi)}{\chi^2}\right)\alpha'+ \left(\frac{2(1-\chi)}{\chi^2}\right)\;.\\
\nonumber&\ldots\ldots&
\end{eqnarray} 
Performing the twist $\alpha=1/\chi$ alone on $\sigma^i\tau^j$ also produces similar rational functions. We notice that the highest power of $\alpha$ in $\mathcal{G}$ is $\mathcal{L}$, as it follows from the fact that $\mathcal{G}$ is a degree-$\mathcal{L}$ polynomial of $\sigma$ and $\tau$. It is easy to see that the action of these operations does not change this degree.  This instructs us to take out a factor $(1-\chi)^{-1}\chi^{-\mathcal{L}}$ from (\ref{scfwiposition}), so that the left side becomes a degree-($\mathcal{L}+1)$ polynomial of $\chi$. Schematically, we can write the new identity as
\begin{equation}\label{identitychi}
f_0+\chi f_1+\chi^2 f_2+\ldots+ \chi^{\mathcal{L}+1}f_{\mathcal{L}+1}=0\;
\end{equation}
where $f_i=f_i(U,V;\alpha')$ are functions of the conformal cross ratios $U$, $V$ and the untwisted R-symmetry variable $\alpha'$. Note that an ambiguity exists in the change of variables (\ref{changeofvariable}), namely, under the exchange of $\chi\leftrightarrow\chi'$ the variables $U$ and $V$ remain the same. Hence by exchanging $\chi$ with $\chi'$ we get from (\ref{identitychi}) another copy of the identity for free 
 \begin{equation}\label{identitychiprime}
f_0+\chi' f_1+\chi'^2 f_2+\ldots+ \chi'^{\mathcal{L}+1}f_{\mathcal{L}+1}=0\;.
\end{equation}
Taking the sum of these two identities, we arrive at the following equation
  \begin{equation}\label{identitysum}
2f_0+(\chi+\chi') f_1+(\chi^2+\chi'^2) f_2+\ldots+(\chi^{\mathcal{L}+1}+\chi'^{\mathcal{L}+1})f_{\mathcal{L}+1}=0\;.
\end{equation}
Crucially, the appearance of $\chi$ and $\chi'$ is now symmetrized and each $\chi^n+\chi'^n$ can be rewritten as a finite linear combination of $U^mV^n$\footnote{This is easy to see by induction: 
\begin{equation}
\chi^n+\chi'^n=(\chi^{n-1}+\chi'^{n-1})(\chi+\chi')-\chi\chi'(\chi^{n-2}+\chi'^{n-2})\;.
\end{equation}
Note $\chi\chi'=U$ and for $n=1$, $\chi+\chi'=U-V+1$.}. After making this replacement, we can now exploit the superconformal Ward identity in Mellin space using elementary manipulations of the inverse Mellin transformation.  

Substituting $\mathcal{G}_{{\rm conn}, LMN}(U,V)$ with its inverse Mellin representation (\ref{inverseGmono}), the following dictionary then becomes clear
\begin{eqnarray}
U\frac{\partial}{\partial U}\quad &\Rightarrow&\quad \; \left[\frac{s}{2}-\frac{\epsilon(k_3+k_4)}{2}+\epsilon\mathcal{L}\right] \times\;,\label{UdU}\\
V\frac{\partial}{\partial V}\quad &\Rightarrow&\quad \; \left[\frac{t}{2}-\frac{\epsilon \min\{k_1+k_4,k_2+k_3\}}{2}\right]\times\;,\label{VdV}\\
U^m V^n\quad &\Rightarrow & \quad\quad\;\; \text{shift $s$ by $-2m$ and $t$ by $-2n$.}\label{actionuv}
\end{eqnarray} 
Note the shifts in the third line act on the {\it reduced} Mellin amplitude, {\it i.e.}, both the Mellin amplitude and the Gamma functions. It becomes more convenient if we preserve in each integrand a common factor of Gamma functions $\Gamma_{k_1k_2k_3k_4}$ as defined in  (\ref{gammafactor}) when we add up the inverse Mellin transformations. The monomial $U^mV^n$ then becomes an operator $\underline{\widehat{U^mV^n}}$\footnote{This operator should not be confused with the operator $\widehat{U^mV^n}$ used in \cite{Rastelli:2017udc,Rastelli:2016nze}. We put an underline to distinguish it.} that only acts on the {\it Mellin amplitude} in the following way
\begin{equation}
U^mV^n\Rightarrow \underline{\widehat{U^mV^n}}
\end{equation}
\begin{equation}\label{UmVn}
\begin{split}
\underline{\widehat{U^mV^n}}\circ \mathcal{M}(s,t)={}&\mathcal{M}(s-2m,t-2n)\left(\frac{\epsilon(k_1+k_2)-s}{2}\right)_m\left(\frac{\epsilon(k_3+k_4)-s}{2}\right)_m\\
{}&\quad\times \left(\frac{\epsilon(k_1+k_4)-t}{2}\right)_n\left(\frac{\epsilon(k_2+k_3)-t}{2}\right)_n\left(\frac{\epsilon(k_1+k_3)-u}{2}\right)_{-m-n}\\
{}&\quad\times \left(\frac{\epsilon(k_2+k_4)-u}{2}\right)_{-m-n}\;.
\end{split}
\end{equation}
Here $(a)_n$ is the Pochhammer symbol. Since (\ref{identitysum}) is a degree-$\mathcal{L}$ polynomial of $\alpha'$, we get in total $\mathcal{L}+1$ identities for the Mellin amplitude $\mathcal{M}(s,t;\sigma,\tau)$. These are the Mellin space superconformal Ward identities.

To close this subsection, let us summarize the procedure for implementing the superconformal Ward identity in Mellin space:
\begin{enumerate}
\item We start with $(\chi\partial_\chi-\epsilon\alpha\partial_\alpha)\mathcal{G}(\chi,\chi';\alpha,\alpha')$. $\mathcal{G}(\chi,\chi';\alpha,\alpha')$ is decomposed into R-symmetry monomials $\sum_{L+M+N=\mathcal{L}}\sigma^M\tau^N\mathcal{G}_{{\rm conn},LMN}(U,V)$ and $\sigma$ and $\tau$ are related to $\alpha$, $\alpha'$ via (\ref{changeofvariable}). We write the action of $\chi\partial_\chi$ as (\ref{chidchi}) and perform the action of $\alpha\partial_\alpha$. The action of $U\partial_U$ and $V\partial_V$ on $\mathcal{G}_{{\rm conn}, LMN}(U,V)$ are not evaluated at this step.
\item We perform the twist $\alpha=1/\chi$ and multiply the expression with $(1-\chi)\chi^{\mathcal{L}}$ to make it a polynomial of $\chi$.
\item We replace all $\chi$ with $\chi'$ and add up the two expressions. All the $\chi$ and $\chi'$ are then rewritten as polynomials of $U$ and $V$.
\item We use the inverse Mellin representations of the correlator. This amounts to replacing each $G_{{\rm conn}, LMN}$ with $\mathcal{M}_{LMN}$. There are additional factors and derivatives of $U$ and $V$. For $U\partial_U$ and $V\partial_V$, we replace them with the  factors (\ref{UdU}) and (\ref{VdV}) that multiply the Mellin amplitude. For monomials $U^mV^n$, we replace them with the operator $\underline{\widehat{U^mV^n}}$ whose action on the Mellin amplitude is given by (\ref{UmVn}).
\item We organize the expression by powers of $\alpha'$. All the polynomial coefficients of $\alpha'$ are linear functions of $\mathcal{M}_{LMN}$ with shifted arguments, and they are required to be zero. These equations are the Mellin space superconformal Ward identities.
\end{enumerate}

\subsection{Bootstrapping Holographic Mellin Amplitudes}\label{bootstrapmellin}

Now we are equipped with the Mellin space superconformal Ward identities, we can formulate another bootstrap-inspired approach which computes holographic correlators entirely within the Mellin space. The idea is straightforward: we formulate an ansatz for the Mellin amplitude and then solve the ansatz using superconformal symmetry. As was reviewed in Section \ref{mellinrep}, the structure of supergravity Mellin amplitudes is very simple. As a function of the Mandelstam variables $s$ and $t$, the Mellin amplitude splits into a singular part and a regular part. The singular part has only simple poles in $s$, $t$ and $u$, whose locations are determined by the spectrum of the theory and the cubic coupling selection rules. At each simple pole, the residue has to be a polynomials of the other independent Mandelstam variable -- it is the linear superposition of Mack polynomials. Because we have in the spectrum particles with maximal spin two, the degree of such a residue polynomial is bounded to be two. The regular part is even simpler. It is a linear polynomial of the Mandelstam variables as is required by the consistency with the flat space limit. Of course each term also depends polynomially on the R-symmetry variables $\sigma$ and $\tau$, with a degree $\mathcal{L}$ determined by the weights $k_i$ of the external operators. Computing the Mellin amplitudes amounts to fixing the coefficients in the residues and in the regular part, and the Mellin space superconformal Ward identities in Section \ref{solscfwi} help us achieve precisely that.

Let us now divide the further description of the ansatz into two scenarios, depending on the number of simple poles in the singular part of the Mellin amplitude is finite or infinite. The former situation occurs in $AdS_5\times S^5$ and $AdS_7\times S^4$. In this case we can use a totally general ansatz: we do not make any specification of the coefficients in the residue polynomials and the regular part, and they are left as unknowns to be solved.  When the four external operators are identical, the Mellin amplitude ansatz is further required to have crossing symmetry. This is the most general ansatz one can write down that is compatible with the qualitative information of the bulk supergravity. Applying and solving the Mellin space superconformal Ward identities are completely straightforward as it is a finite problem. For the latter scenario, such as in $AdS_4\times S^7$ , proceeding with such a generic ansatz appears to be technically involved because we need infinitely many coefficients to parameterize the ansatz. We simplify the problem by using an ansatz parallel to the ones used in the position space method. To be precise, we will use the explicit Mellin amplitudes of the exchange Witten diagrams (but only the singular part) and write the singular part of the ansatz as a linear combination of such exchange Mellin amplitudes. These amplitudes are not hard to obtain because it is known that the Mellin amplitude of a conformal block with the same quantum numbers of the exchanged single-trace operator has the same pole and same residues. The Mellin amplitude of conformal blocks can be found in, {\it e.g.}, \cite{Mack:2009mi,Fitzpatrick:2011hu}. For the regular part, we will use the same general parameterization as in the former case. Such an ansatz is general enough to encompass the correct answer, but does not have as much power to exclude other regions in the space of ansatz as it did for $AdS_5\times S^5$ and $AdS_7\times S^4$.

In the following, we will demonstrate the method with the general ansatz for the case of $AdS_5\times S^5$ stress-tensor multiplet four-point function. 

\subsubsection{An example: stress-tensor multiplet four-point function for $AdS_5\times S^5$}
We reproduce in this subsection the four-point function $\langle\mathcal{O}_2\mathcal{O}_2\mathcal{O}_2\mathcal{O}_2\rangle$ for IIB supergravity on $AdS_5\times S^5$, dual to 4d $\mathcal{N}=4$ SYM in the large $N$, infinite 't Hooft coupling limit. The one-half BPS operator $\mathcal{O}_2$ has dimension $\Delta=2$ and $SU(4)$ Dynkin label $[0,2,0]$, and is the superconformal primary of the stress-tensor multiplet. In the bulk supergravity, $\mathcal{O}_2$ is dual to a scalar field with squared mass $m^2=-4$. Using the selection rules in Section 2 of \cite{Rastelli:2017udc}, we find that the only contributing fields in the exchange Witten diagrams are: the scalar field itself, a vector field with $\Delta=3$ and a graviton field with $\Delta=4$. All these fields have coinciding conformal twist $\tau=2$. The simple poles in their Mellin amplitudes should truncate to just a single one at 2. Hence the singular part of the ansatz should consist of the following terms
\begin{equation}
\mathcal{M}_s+\mathcal{M}_t+\mathcal{M}_u
\end{equation}
where
\begin{eqnarray}
\mathcal{M}_s&=&\sum_{{\fontsize{4}{3}\selectfont\begin{split}{}&0\leq i,j\leq 2,\\{}&0\leq i+j\leq 2\end{split}}}\sum_{{\fontsize{4}{3}\selectfont0\leq a\leq 2}}\frac{\lambda_{ij;a}^{(s)}\;\sigma^i\tau^jt^a}{s-2}\;,\\
\mathcal{M}_t&=&\sum_{{\fontsize{4}{3}\selectfont\begin{split}{}&0\leq i,j\leq 2,\\{}&0\leq i+j\leq 2\end{split}}}\sum_{{\fontsize{4}{3}\selectfont0\leq a\leq 2}}\frac{\lambda_{ij;a}^{(t)}\;\sigma^i\tau^ju^a}{t-2}\;,\\
\mathcal{M}_u&=&\sum_{{\fontsize{4}{3}\selectfont\begin{split}{}&0\leq i,j\leq 2,\\{}&0\leq i+j\leq 2\end{split}}}\sum_{{\fontsize{4}{3}\selectfont0\leq a\leq 2}}\frac{\lambda_{ij;a}^{(u)}\;\sigma^i\tau^js^a}{u-2}\;,
\end{eqnarray}
and $s+t+u=8$. We should also add the following polynomial term that represents the contact interactions
\begin{equation}
\mathcal{M}_{c}=\sum_{{\fontsize{4}{3}\selectfont\begin{split}{}&0\leq i,j\leq 2,\\{}&0\leq i+j\leq 2\end{split}}}\sum_{{\fontsize{4}{3}\selectfont\begin{split}{}&0\leq a,b\leq 1,\\{}&0\leq a+b\leq 1\end{split}}}\mu_{ij;ab}\;\sigma^i\tau^j s^a t^b\;.
\end{equation}
The most general ansatz in the supergravity result is therefore
\begin{equation}
\mathcal{M}_{\text{ansatz}}=\mathcal{M}_s+\mathcal{M}_t+\mathcal{M}_u+\mathcal{M}_c\;,
\end{equation} 
 and it should satisfy  in addition the following crossing equations
 \begin{equation}
 \mathcal{M}_{\text{ansatz}}(s,t;\sigma,\tau)=\tau^2\mathcal{M}_{\text{ansatz}}(t,s;\sigma/\tau,1/\tau)=\sigma^2\mathcal{M}_{\text{ansatz}}(u,t;1/\sigma,\tau/\sigma)\;
 \end{equation} 
where $u=8-s-t$.
 
We now impose the Mellin space superconformal Ward identities following the procedure given in Section \ref{solscfwi}. We obtain three sets of equations of $\mathcal{M}_{\text ansatz}$ with shifted arguments, as (\ref{identitysum}) is degree-two in $\alpha'$ and the coefficients of $\alpha'^n$ must separately vanish. Bringing all terms in each equation to the minimal common denominator, the numerator is a polynomial in $s$ and $t$ with coefficients linearly depending on the unfixed parameters. Requiring these coefficients to vanish gives us a set of linear equations.

We solve these equations together with the crossing equations and we arrive at the following solution
\begin{eqnarray}
\mathcal{M}&=& C\frac{\sigma(4u-4t+8)+\tau(4t-4u+8)+(u^2+t^2-6u-6t+16)}{s-2}\\
&&+C\frac{\tau(\sigma(4u-4s+8)+\tau(s^2+u^2-6s-6u+16)+(4s-4u+8))}{t-2}\\
&&+C\frac{\sigma(\sigma(s^2+t^2-6s-6t+16)+\tau(4t-4s+8)+(4s-4t+8))}{u-2}\\
&&+C(-s-u\sigma^2-t\tau^2+4(t+u-2)\sigma\tau+4(s+u-2)\sigma+4(s+t-2)\tau)
\end{eqnarray}
where $C$ is an unfixed overall coefficient. This answer agrees with the original supergravity result \cite{Arutyunov:2000py}.

\section{Part II: Applications}\label{partII}
We now turn to the applications and obtain new results.\footnote{In this paper we focus on  tree-level  applications. The method of imposing superconformal constraints on Mellin amplitudes works at the loop level as well. Recently much progress has been made on studying 4d $\mathcal{N}=4$ SYM in the supergravity limit at one loop (see, {\it e.g.}, \cite{Aprile:2017bgs,Aprile:2017xsp,Alday:2017xua,Alday:2017vkk,Aprile:2017qoy}). It would be interesting to reexamine it in Mellin space using our technique. The universality of our technique might also make it convenient to extend the analysis to other duality pairs in different dimensions.} In Section \ref{6dapplication}, we study the next-next-to-extremal four-point functions for the 6d (2,0) theories. We will use a general ansatz for the Mellin amplitude such as was used for the 4d $\mathcal{N}=4$ SYM in the previous section. Solving the Mellin space superconformal Ward identity fixes the answer uniquely up to an overall constant. The Mellin amplitude of this family of four-point functions can be further cast into a succinct expression with only one term, using a difference operator that we will introduce in Section \ref{scfkinematics}. In Section \ref{3dapplication} we turn to computing the holographic four-point function of the stress-tensor multiplet for 3d $\mathcal{N}=8$ SCFTs. In this case we encounter a new technical difficulty, namely, that there necessarily exists infinite series of simple poles in the Mellin amplitude. Despite the presence of infinitely many poles, the Mellin space superconformal Ward identity can be solved and the final result is very simple. We fix the overall constant and extract the anomalous dimension for the R-symmetry singlet double-trace operator with the lowest conformal dimension in Section \ref{anom}. The anomalous dimension is found to be compatible with the numerical bootstrap result \cite{Chester:2014fya,Agmon:2017xes}. 

\subsection{6D (2,0): 11D Supergravity on $AdS_7\times S^4$}\label{6dapplication}
In this section we apply the Mellin space method to compute next-next-to-extremal correlators for the eleven dimensional supergravity compactified on $AdS_7\times S^4$. The supergravity theory is conjectured to be dual to an superconformal field theory in six dimensions with (2,0) superconformal symmetry. The SCFT is the low energy effective theory of $N$ coincident $M5$-branes, with $N$ large. The theory has a superconformal group $OSp(8^*|4)$, which contains as bosonic subgroups the conformal symmetry group $SO(6,2)$ and R-symmetry group $USp(4)=SO(5)$. The one-half BPS operators $\mathcal{O}_k^{I_1\ldots I_k}$ transform in the rank-$k$ symmetric-traceless representation of the R-symmetry group $SO(5)$ and has conformal dimension $\Delta=2k$. These operators are dual to the Kaluza-Klein modes, labelled by $s_k$,  of a scalar field in the bulk. They have squared mass $m^2=2k(2k-6)$ and the rank-$k$ symmetric-traceless representation is captured by their harmonics on $S^4$. It is required that $k\geq 2$.  

The {\it extremality} of a four-point function is defined by 
\begin{equation}
E=k_2+k_3+k_4-k_1\;
\end{equation}
where the ordering of the weights $k_1\geq k_2\geq k_3\geq k_4$ is assumed. R-symmetry selection rules determine that $E$ is an even integer. When $E=0,2$, the four-point functions are respectively said to be {\it extremal} and {\it next-to-extremal}. For $\mathcal{N}=4$ SYM in 4D, such correlators are protected by non-renormalization theorems \cite{DHoker:1999jke,Bianchi:1999ie,Eden:1999kw,Erdmenger:1999pz,Eden:2000gg} and they have the same value as in the free limit of the theory. For 6D (2,0) theories, the notion of ``non-renormalization'' is moot due to the absence of an exactly marginal coupling.  However, one can still argue from the finiteness of the boundary correlator that the supergravity couplings should vanish as the relevant Witten diagrams are divergent \cite{DHoker:2000pvz}. A regularization procedure needs to be taken, and the end result is that the correlators computed from supergravity are rational functions of the cross ratios -- the dependence on the cross ratios is similar to that in a generalized free field theory. The $E=4$ case is the first case free of such subtleties and the four-point function starts to depend on the cross ratios in a more non-trivial manner. Such {\it next-next-to-extremal} correlators will be the focus of this section. We will take two operators with $k_3=k_4=k+2$ and the other two operators with $k_1=n+k$, $k_2=n-k$. The same class of four-point functions have been studied for IIB $AdS_5\times S^5$ supergravity in \cite{Uruchurtu:2008kp} and we will find that the solution for $AdS_7\times S^4$ takes a very similar form.

The plan of this section is as follows. We start with some comments on the superconformal kinematics in Section \ref{scfkinematics}. We show in particular the Mellin amplitude can be packaged into a simpler auxiliary amplitude with the help of a difference operator. In Section \ref{summarysugra} we review the selection rules of 11D Supergravity compactified on $AdS_7\times S^4$ which are needed to formulate the Mellin amplitude ansatz. Finally, we solve the next-next-to-extremal four-point functions and find an extremely simple answer. 

\subsubsection{Some Superconformal Kinematics}\label{scfkinematics}
In this subsection we extend the discussion in \cite{longads7}
 about the superconformal kinematics of four-point functions in 6d (2,0) theories to the general case where the four operators have unequal weights. The total Mellin amplitude $\mathcal{M}$ is  captured by an auxiliary amplitude $\widetilde{\mathcal{M}}$, and the two are related by a difference operator $\widehat{\Theta}$. The derivation of this operator is a straightforward generalization of \cite{longads7}. We will refer the readers to \cite{longads7} for the technical details.
 
 Let us begin with the position space solution of the superconformal Ward identity (\ref{scfwiposition}). In this subsection, $\epsilon$ is specified to be 2. We consider first a special situation where the R-symmetry cross ratios of the four-point function are restricted to a special slice $\alpha=\alpha'=1/\chi$. In this special kinematic configuration the superconformal Ward identity (\ref{scfwiposition}) reduces to the following equation
\begin{equation}
\chi\partial_{\chi}\mathcal{G}(\chi,\chi';1/\chi,1/\chi)=0\;,
\end{equation}
whose solution is simply
\begin{equation}
\mathcal{G}(\chi,\chi';1/\chi,1/\chi)=f(\chi')\;.
\end{equation}
The holomorphic function $f(\chi')$ can be interpreted as the four-point function of a chiral CFT, obtained from a construction known as the chiral algebra twist \cite{Beem:2013sza,Beem:2014kka,Beem:2014rza}.\footnote{A similar construction also exists in 3d where operators are inserted on a line and resulting theory is topological ({\it i.e.}, depending only on the ordering on the line) \cite{Chester:2014mea,Beem:2016cbd}.} This holomorphic correlator is completely determined by the shortened multiplets of the theory.

 The superconformal Ward identity in 6D was also solved in full generality, and the four-point correlator  was shown to be a sum of two parts \cite{Dolan:2004mu}:
 \begin{equation}\label{scfwisol}
\mathcal{G}(U,V;\sigma,\tau)=\mathcal{F}(U,V;\sigma,\tau)+\mathcal{K}(U,V;\sigma,\tau)\;.
\end{equation}
Here $\mathcal{F}(U,V;\sigma,\tau)$ is an ``inhomogeneous solution'' to the superconformal Ward identity, in the sense that it gives the holomorphic correlator under the above twist $\alpha=\alpha'=1/\chi'$. On the other hand, $\mathcal{K}(U,V;\sigma,\tau)$ is a ``homogeneous solution'' and vanishes when twisted. $\mathcal{K}(U,V;\sigma,\tau)$ is further written as a differential operator $\Upsilon$ acting on a unconstrained function $\mathcal{H}(U,V;\sigma,\tau)$ of degree $\mathcal{L}-2$ in $\sigma$ and $\tau$
\begin{equation}
\mathcal{K}(U,V;\sigma,\tau)=\Upsilon\circ \mathcal{H}(U,V;\sigma,\tau)\;.
\end{equation}
The correct expression for $\Upsilon$ is given in \cite{longads7} but is rather involved and will not be presented here. 

The above solution of the superconformal Ward identity has a counterpart in Mellin space. To obtain it we simply write the position space solution (\ref{scfwisol}) as an identity of inverse Mellin transformation. We will only take the connected part of correlator. On the left side, we have as in (\ref{inverseG})
\begin{equation}\label{inverseG6d}
 \mathcal{G}_{\rm conn}(U,V;\sigma,\tau) = \int \frac{ds}{2} \frac{dt}{2}\; U^{\frac{s}{2}-(k_3+k_4-2\mathcal{L})}V^{\frac{t}{2}-\min\{k_1+k_4,k_2+k_3\}}\mathcal{M}(s,t;\sigma,\tau) \Gamma_{k_1k_2k_3k_4}\;.
\end{equation}
On the right side, we can argue that we should define the Mellin amplitude of $\mathcal{F}$ to be {\it zero}, using the argument of \cite{Rastelli:2017udc}.\footnote{Setting the Mellin amplitude of $\mathcal{F}$ to zero however does not imply that the information of $\mathcal{F}$ is lost. In the inverse Mellin representation of the correlator, $\mathcal{F}$ can still be reproduced through a ``domain pinching mechanism'' by carefully keeping track of the integration contours. For more details, see \cite{Rastelli:2017udc}.} For $\mathcal{H}$, we are instructed to use the following inverse Mellin transformation
\begin{equation}\label{inverseH}
\mathcal{H}(U,V;\sigma,\tau) = \int \frac{ds}{2} \frac{dt}{2} U^{\frac{s}{2}-k_3-k_4+2{\mathcal{L}}+1}V^{\frac{t}{2}-{\min\{k_1+k_4,k_2+k_3\}+1}}\widetilde{\mathcal{M}}(s,t;\sigma,\tau) \tilde{\Gamma}_{k_1k_2k_3k_4}
\end{equation}
where
\begin{equation}
\begin{split}
\tilde{\Gamma}_{k_1k_2k_3k_4} \equiv {}&\Gamma[-\frac{s}{2}+(k_1+k_2)]\Gamma[-\frac{s}{2}+k_3+k_4)]\Gamma[-\frac{t}{2}+(k_2+k_3)]\\
\times{}&\Gamma[-\frac{t}{2}+(k_1+k_4)]\Gamma[-\frac{\tilde{u}}{2}+(k_1+k_3)]\Gamma[-\frac{\tilde{u}}{2}+(k_2+k_4)]\; , \\
\tilde{u} \equiv & \;2(k_1+k_2+k_3+k_4)-s-t-6\;.
\end{split}
\end{equation}
The motivation for such a choice of parameters inside the transformation is to manifest the Bose symmetry of $\widetilde{\mathcal{M}}$ ({\it c.f.} \cite{longads7} for a detailed discussion). It is also useful to decompose $\widetilde{\mathcal{M}}$ into R-symmetry monomials and obtain the R-symmetry partial amplitudes $\widetilde{\mathcal{M}}_{lmn}$
\begin{equation}
\widetilde{\mathcal{M}}(s,t;\sigma,\tau)=\sum_{l+m+n=\mathcal{L}-2}\sigma^m\tau^n\widetilde{\mathcal{M}}_{lmn}(s,t)\;.
\end{equation}

We are now ready to write down the Mellin space version of the solution (\ref{scfwisol}), which takes the simple form
\begin{equation}
\mathcal{M}(s,t;\sigma,\tau)=\widehat{\Theta}\circ \widetilde{\mathcal{M}}(s,t;\sigma,\tau)\;.
\end{equation}
Here $\widehat\Theta$ is a difference operator that acts on the auxiliary amplitude $\widetilde{\mathcal{M}}$ 
\begin{equation} \label{Theta}
\widehat\Theta=-\frac{1}{4}\big((XY)\widehat{B\mathfrak{R}}+(XZ)\widehat{C\mathfrak{R}}+(YZ)\widehat{A\mathfrak{R}}\big)\;,
\end{equation}
with $\mathfrak{R}$ given by
\begin{equation}
\mathfrak{R}= B C+A \sigma^2 C+A B \tau^2+\sigma C (-A-B+C)+ \tau B (-A+B-C)+\sigma\tau A (A-B-C)\;.
\end{equation}
The operators $\widehat{A\mathfrak{R}}$, $\widehat{B\mathfrak{R}}$ and $\widehat{C\mathfrak{R}}$ are written schematically, and should be understood as first expanding $A\mathfrak{R}$, $B\mathfrak{R}$, $C\mathfrak{R}$ into monomials\footnote{The meaning of the letters $A$, $B$, $C$ is explained in \cite{longads7}. However knowing their meaning is immaterial to understand the prescription, and one can abstractly think of them as devices bookkeeping the three powers $(\alpha,3-\alpha-\beta,\beta)$ that enter (\ref{monomial}).} of the form $A^\alpha B^{3-\alpha-\beta}C^\beta$ and then interpreting each monomial as a difference operator acting on a $\widetilde{\mathcal{M}}_{lmn}(s,t)$.  The explicit action of a monomial is given as follows 
\begin{equation}\label{monomial}
\begin{split}
\widehat{A^\alpha B^{3-\alpha-\beta}C^\beta} {}&\circ  \widetilde{\mathcal{M}}_{lmn}(s,t)\equiv  \widetilde{\mathcal{M}}_{lmn}(s-2\alpha,t-2\beta)\times \left(\frac{2(k_1+k_2)-s}{2}\right)_\alpha \\
{}&\times \left(\frac{2(k_3+k_4)-s}{2}\right)_\alpha \left(\frac{2(k_1+k_4)-t}{2}\right)_\beta \left(\frac{2(k_2+k_3)-t}{2}\right)_\beta\\
{}&\times \left(\frac{2(k_1+k_3)-u}{2}\right)_{3-\alpha-\beta} \left(\frac{2(k_2+k_4)-u}{2}\right)_{3-\alpha-\beta}\;.
\end{split}
\end{equation}
Finally, the multiplicative factors made of $X$, $Y$ and $Z$ in (\ref{Theta}) depend on the value of $(l,m,n)$ of each $\widetilde{\mathcal{M}}_{lmn}(s,t)$ on which the associated difference operator is acting. They take the following values
\begin{equation}
\begin{split}
X={}&s+4l+2-2\min\{k_1+k_2,k_3+k_4\}\;,\\
Y={}&t+4n+2-2\min\{k_1+k_4,k_2+k_3\}\;,\\
Z={}&u+4m+2-2\min\{k_1+k_3,k_2+k_4\}\;.
\end{split}
\end{equation}
The above operator $\widehat{\Theta}$ can be obtained by acting the differential operator $\Upsilon$ on (\ref{inverseH}) and then interpreting the monomials of cross ratios as difference operators. This procedure is detailed in \cite{longads7}. We will not give the derivation here.

\subsubsection{Selection Rules}\label{summarysugra}
We now proceed to summarizing the requisite qualitative information for formulating the Mellin space ansatz. We focus on the supergravity cubic vertices, for which the only information we need are the selection rules. An obvious constraint comes from the following tensor product rules of R-symmetry representations\footnote{In this section, we are using the $USp(4)$ Dynkin labels.}
\begin{equation}
{[0,p]}\otimes {[0,q]}= \sum_{a=0}^{\frac{p+q-|p-q|}{2}}\sum_{b=0}^a[2(a-b),2b+|p-q|]\;.
\end{equation}
On the left hand side, $[0,p]$ and $[0,q]$ are the irreducible representations carried by fields $s_p$ and $s_q$ which are two external legs joined at a cubic vertex. On the right hand side, the sum defines a set inside which the irreducible representation of an exchanged field must fall. We collect in Table \ref{ads7spectrum} the list of bulk fields $\{\phi_{\mu_1\ldots\mu_\ell}\}$ {\it a priori} allowed in an exchange diagram with four external fields $s_{k_i}$ if only R-symmetry selection rules are imposed.

\begin{table}[t]
\begin{center}\begin{tabular}{|c|c|c|c|c|c|c|c|c|c|}\hline field & $\ell$ &  R-irrep & $m^2$ & $\Delta$ & $k=2$ & $k=3$ & $k=4$ & $k=5$ & $k=6$ \\ \hline$\varphi_{\mu\nu,k}$ & 2 & $[0,k-2]$ & $4(k-2)(k+1)$ & $2k+2$ & {6} & 8 & {10} & 12 & {14} \\$A_{\mu,k}$ & 1 & $[2,k-2]$ & $4k(k-2)$ & $2k+1$ & {5} & 7 & {9} & 11 & {13} \\$C_{\mu,k}$ & 1 & $[2,k-4]$ & $4(k-1)(k+1)$ & $2k+3$ & - & - & {11} & 13 & {15} \\$s_{k}$ & 0 & $[0,k]$ & $4k(k-3)$ & $2k$ & {4} & 6 & {8} & 10 & {12} \\$t_{k}$ & 0 & $[0,k-4]$ & $4(k-1)(k+2)$ & $2k+4$ & - & - & {12} & 14 & 16 \\$r_{k}$ & 0 & $[4,k-4]$ & $4(k-2)(k+1)$ & $2k+2$ & - & - & {10} & 12 & {14}\\\hline\end{tabular} \caption{\small KK modes contributing to exchange diagrams allowed by R-symmetry selection rules.}\label{ads7spectrum}
\end{center}
\end{table}

Furthermore, for the cubic coupling to be non-vanishing, the following additional condition on the twist of the exchanged field
\begin{equation}\label{twistselect}
\Delta-\ell< \Delta_1+\Delta_2=2k_1+2k_2
\end{equation} 
must be satisfied. This condition follows from some simple reasoning. In order for the cubic vertex $s_{k_1}s_{k_2}\phi_{\mu_1\ldots\mu_\ell}$ to be non-zero, the vertex $s_{k_1}s_{k_2}s_{k_3}$ must be non-zero, where $s_{k_3}$ is the superprimary of the field $\phi_{\mu_1\ldots\mu_\ell}$. It is easy to check that the twists of all descendants  have the same parity as the primary, {\it i.e.}, $\tau_{s}-\tau_{\phi}\in 2\mathbb{Z}$. R-symmetry requires that the CFT correlator $\langle \mathcal{O}_{k_1}\mathcal{O}_{k_2}\mathcal{O}_{k_3}\rangle$ is zero if $k_3> k_1+k_2$ and therefore the corresponding cubic vertex $s_{k_1}s_{k_2}s_{k_3}$ vanishes. The extremal case of $k_3=k_1+k_2$ is more subtle because the three-point function $\langle \mathcal{O}_{k_1}\mathcal{O}_{k_2}\mathcal{O}_{k_3=k_1+k_2}\rangle$ from the boundary CFT  is usually nonzero. One way to reproduce $\langle \mathcal{O}_{k_1}\mathcal{O}_{k_2}\mathcal{O}_{k_3=k_1+k_2}\rangle$ from holography is by analytic continuation in $k_3$ \cite{Lee:1998bxa,DHoker:1999jke}. The cubic coupling should behave like $c_{k_1k_2k_3}\sim (k_3-k_2-k_1)$ as the three-point Witten diagram diverges as $1/(k_3-k_2-k_1)$. This makes the product of the two yield the finite correct answer. From this point of view, the extremal coupling $c_{k_1k_2k_1+k_2}$ is necessary to vanish to maintain the finiteness of the three-point function. This twist selection rule also admits a natural explanation in Mellin space, see Section 3.2 and 3.3 of \cite{Rastelli:2017udc} for a detailed discussion.

Now let us specialize our discussion to the next-next-to-extremal four-point functions. For now, we are going to assume $n\geq 2k+2$ so that $k_1=n+k$, $k_2=n-k$, $k_3=k+2$, $k_4=k+2$ is consistent with the previously defined ordering $k_1\geq k_2\geq k_3\geq k_4$. The possibility of $k+2\leq n<2k+2$ just amounts to relabelling the last three operators. In the s-channel ($s_{k_1}s_{k_2}\rightarrow s_{k_3}s_{k_4}$), we find the following common irreducible representations in the tensor products of ${[0,n+k]}\otimes{[0,n-k]}$ and $[0,k+2]\otimes[0,k+2]$
\begin{equation}
[0,2k+4]\;,\quad [2,2k+2]\;,\quad [4,2k]\;,\quad [0,2k+2]\;,\quad [2,2k]\;,\quad [0,2k]\;.
\end{equation}
By comparing with Table \ref{ads7spectrum} and using the twist selection rule (\ref{twistselect}), we find that the only fields exchanged in the s-channel are the scalar field $s_{2k+2}$, the vector field $A_{\mu,2k+2}$ and the massive symmetric tensor field $\varphi_{\mu\nu,2k+2}$.

Similarly in the t-channel ($s_{k_1}s_{k_4}\rightarrow s_{k_2}s_{k_3}$) and u-channel ($s_{k_1}s_{k_3}\rightarrow s_{k_2}s_{k_4}$), we find the following common irreducible representations of R-symmetry 
\begin{equation}
[0,n+2]\;,\quad [2,n]\;,\quad [4,n-2]\;,\quad [0,n]\;,\quad [2,n-2]\;,\quad [0,n-2]\;.
\end{equation}
Table \ref{ads7spectrum}  and the twist selection rule (\ref{twistselect}) then dictate that only fields $s_{2n}$, $A_{\mu,2n}$ and $\varphi_{\mu\nu,2n}$ can be exchanged in the t and u-channel.

\subsubsection{Next-Next-to-Extremal Four-Point Functions}\label{nearextremal}
We are now ready to write down an ansatz for next-next-to-extremal correlators for $AdS_7\times S^4$.  In Section \ref{summarysugra}, we have identified the fields that can be exchanged in each channel. This allows us to predict the position of poles using the knowledge of the Mellin amplitude of exchange diagrams reviewed in Section \ref{mellinrep}. More precisely, the s-channel exchange diagrams involve three fields with the same conformal twist $\tau=4k+4$. The Mellin amplitude should therefore have a leading simple pole at $s=4k+4$ and a satellite pole at $s=4k+6$. Moreover, since the maximal spin of the exchanged fields is two, we anticipate that the residues at these simple poles are degree-two polynomials in the other Mandelstam variable $t$. The following terms therefore should be part of the ansatz for the total Mellin amplitude
\begin{equation}
\mathcal{M}_s(s,t;\sigma,\tau)=\sum_{{\fontsize{4}{3}\selectfont\begin{split}{}&0\leq i,j\leq 2,\\{}&0\leq i+j\leq 2\end{split}}}\sum_{{\fontsize{4}{3}\selectfont0\leq a\leq 2}}\frac{\lambda_{ij;a}^{(s,1)}\;\sigma^i\tau^jt^a}{s-(4k+4)}+\sum_{{\fontsize{4}{3}\selectfont\begin{split}{}&0\leq i,j\leq 2,\\{}&0\leq i+j\leq 2\end{split}}}\sum_{{\fontsize{4}{3}\selectfont0\leq a\leq 2}}\frac{\lambda_{ij;a}^{(s,2)}\;\sigma^i\tau^jt^a}{s-(4k+6)}\;.
\end{equation}
Notice that in the residues in the above ansatz we have left the dependence on the R-symmetry variables and $t$  completely arbitrary, apart from the bounded degrees.

Similarly, for t and u-channel, we also have three types of field exchanges with the same conformal twist $\tau=2n$ and maximal spin two. We expect simple poles at $t=2n$, $t=2n+2$, $u=2n$ and $u=2n+2$. The residues at these simple poles are also degree-two polynomials. We therefore further include the following $\mathcal{M}_t$ and $\mathcal{M}_u$
 \begin{equation}
\mathcal{M}_t(s,t;\sigma,\tau)=\sum_{{\fontsize{4}{3}\selectfont\begin{split}{}&0\leq i,j\leq 2,\\{}&0\leq i+j\leq 2\end{split}}}\sum_{{\fontsize{4}{3}\selectfont0\leq a\leq 2}}\frac{\lambda_{ij;a}^{(t,1)}\;\sigma^i\tau^ju^a}{t-2n}+\sum_{{\fontsize{4}{3}\selectfont\begin{split}{}&0\leq i,j\leq 2,\\{}&0\leq i+j\leq 2\end{split}}}\sum_{{\fontsize{4}{3}\selectfont0\leq a\leq 2}}\frac{\lambda_{ij;a}^{(t,2)}\;\sigma^i\tau^ju^a}{t-(2n+2)}\;,
\end{equation}
\begin{equation}
\mathcal{M}_u(s,t;\sigma,\tau)=\sum_{{\fontsize{4}{3}\selectfont\begin{split}{}&0\leq i,j\leq 2,\\{}&0\leq i+j\leq 2\end{split}}}\sum_{{\fontsize{4}{3}\selectfont0\leq a\leq 2}}\frac{\lambda_{ij;a}^{(u,1)}\;\sigma^i\tau^js^a}{u-2n}+\sum_{{\fontsize{4}{3}\selectfont\begin{split}{}&0\leq i,j\leq 2,\\{}&0\leq i+j\leq 2\end{split}}}\sum_{{\fontsize{4}{3}\selectfont0\leq a\leq 2}}\frac{\lambda_{ij;a}^{(u,2)}\;\sigma^i\tau^js^a}{u-(2n+2)}\;,
\end{equation}
with $u=4(n+k+2)-s-t$.

This has exhausted all the poles in the Mellin amplitude. Additionally we should also have the following polynomial piece to account for the contact interactions
\begin{equation}
\mathcal{M}_c(s,t;\sigma,\tau)=\sum_{{\fontsize{4}{3}\selectfont\begin{split}{}&0\leq i,j\leq 2,\\{}&0\leq i+j\leq 2\end{split}}}\sum_{\fontsize{4}{3}\selectfont\begin{split}{}&0\leq a,b\leq 1,\\{}&0\leq a+b\leq 1\end{split}}\mu_{ij;ab}\;\sigma^i\tau^js^at^b\;.
\end{equation}
The full general ansatz is then the sum of the four parts
\begin{equation}
\mathcal{M}_{\text{ansatz}}=\mathcal{M}_s+\mathcal{M}_t+\mathcal{M}_u+\mathcal{M}_c\;.
\end{equation} 

Now we impose the Mellin space superconformal Ward identity. Note $\mathcal{M}_{\text ansatz}$ has finitely many terms and contains only finitely many unknown parameters. The problem is therefore completely finite and elementary. Solving these constraints is very straightforward and we find the answer is unique up to an overall constant. For arbitrary $n$ and $k$, we find the following solution
\begin{equation}
\mathcal{M}(s,t;\sigma,\tau)=\mathcal{M}_1(s,t)+\sigma^2\mathcal{M}_{\sigma^2}(s,t)+\tau^2\mathcal{M}_{\tau^2}(s,t)+\sigma\mathcal{M}_{\sigma}(s,t)+\tau\mathcal{M}_{\tau}(s,t)+\sigma\tau \mathcal{M}_{\sigma\tau}(s,t)\;,
\end{equation}
with the R-symmetry partial amplitudes given by 
\begin{equation}\footnotesize
\begin{split}
\mathcal{M}_1(s,t)&=-C_{n,k}\frac{ (k+1) (4 k+2 n-t+4) (4 k-2 n+t)}{8(s-4k-4)}-\frac{C_{n,k}}{32} (2 k+2 n+1) (4 k+2 n-t+4)\\
&+C_{n,k}\frac{(2 k-2 n+3) (4 k+2 n-t+4) (4 k-2 n+t+2)}{32(s-4k-6)}\;,\\
\mathcal{M}_{\sigma^2}(s,t)&=C_{n,k}\frac{n (4 k+2 n-t+4) (-4 k+2 n+t-8)}{16(s+t-4k-2n-8)}-\frac{C_{n,k}}{32} (2 k+2 n+1) (4 k+2 n-t+4)\\
&-C_{n,k}\frac{(2 k+1) (4 k-2 n-t+6) (4 k+2 n-t+4)}{32(s+t-4k-2n-6)}\;,\\
\mathcal{M}_{\tau^2}(s,t)&=-C_{n,k}\frac{n (s-4) (4 n-s)}{16(t-2n)}-C_{n,k}\frac{(2 k+1) (s-2) (4 n-s)}{32(t-2n-2)}-\frac{C_{n,k}}{32} (2 k+2 n+1) (4 n-s)\;,
\end{split}
\end{equation}
\begin{equation}\footnotesize
\begin{split}
\mathcal{M}_{\sigma}(s,t)=&C_{n,k}\frac{(k+1) (4 k+3) (4 k+2 n-t+4)}{4(s-4k-4)}-C_{n,k}\frac{(k+1) (2 k-2 n+3) (4 k+2 n-t+4)}{4(s-4k-6)}\\
&-C_{n,k}\frac{(2 k+1) n (4 k+2 n-t+4)}{8(s+t-4k-2n-6)}-C_{n,k}\frac{n (2 n-1) (4 k+2 n-t+4)}{8(s+t-4k-2n-8)}\\
&+\frac{C_{n,k}}{16} (2 k+2 n+1) (4 k+2 n-t+4)\;,\\
\mathcal{M}_{\tau}(s,t)=&C_{n,k}\frac{(k+1) (4 k+3) (4 k-2 n+t)}{4(s-4k-4)}-C_{n,k}\frac{(k+1) (2 k-2 n+3) (4 k-2 n+t+2)}{4(s-4k-6)}\\
&+C_{n,k}\frac{n (2 n-1) (s-4)}{8(t-2n)}+C_{n,k}\frac{(2 k+1) n (s-2)}{8(t-2n-2)}+\frac{C_{n,k}}{16} (2 (k+n) (4 k+s+t)-2 n+s+t-4)\\
\mathcal{M}_{\sigma\tau}(s,t)=&C_{n,k}\frac{n (2 n-1) (4 n-s)}{8(t-2n)}+C_{n,k}\frac{(2 k+1) n (4 n-s)}{8(t-2n-2)}-C_{n,k}\frac{(2 k+1) n (-4 k+2 n+t-6)}{8(s+t-4k-2n-6)}\\
&-C_{n,k}\frac{n (2 n-1) (-4 k+2 n+t-8)}{8(s+t-4k-2n-8)}+\frac{C_{n,k}}{16} (-2 s (k+n)+4 n (3 k+3 n+1)-s)\;.
\end{split}
\end{equation}
The number $C_{n,k}$ above is an overall normalization that depends on the value of $n$, $k$ and cannot be fixed by the symmetry considerations alone. This remaining parameter can be however determined by using, {\it e.g.}, the three-point function of the three scalar fields. 

The next-next-to-extremal four-point Mellin amplitude has a hidden simplicity. Using the prescription from Section \ref{scfkinematics}, we find that the next-next-to-extremal Mellin amplitude can be written into the following remarkably simple one-term expression in terms of the auxiliary Mellin amplitude
\begin{equation}
\widetilde{\mathcal{M}}(s,t)=\frac{C_{n,k}}{(s-4k-6)(s-4k-4)(t-2n-2)(t-2n)(\tilde{u}-2n-2)(\tilde{u}-2n)}\;.
\end{equation}

Translating the above result into position space, we find $\mathcal{H}$ consists of only one single $\bar{D}$-function (see (5.4) and (3.30) of \cite{Rastelli:2017udc} for our convention of $\bar{D}$-function and its Mellin transformation)
\begin{equation}
\mathcal{H}=64\;C_{n,k}U^{2n-2k+1}V^{-1}\bar{D}_{2n+2k+3,2n-2k-1,2k+3,2k+3}\;.
\end{equation}
This is very similar to the $AdS_5\times S^5$ case \cite{Uruchurtu:2011wh}, for which the next-next-to-extremal four-point functions\footnote{The full solution to the superconformal Ward identity can still be written in the form of $\mathcal{F}+\mathcal{K}$, where $\mathcal{H}$ enters $\mathcal{K}$ as $\mathcal{K}=(1-\chi\alpha)(1-\chi\alpha')(1-\chi'\alpha)(1-\chi'\alpha')\mathcal{H}$.} have $\mathcal{H}=\tilde{C}_{n,k}U^{n-k}V^{-1}\bar{D}_{n+k+2,n-k,k+2,k+2}$ and $\tilde{C}_{n,k}$ is some constant.

When we take the special value of $n=2$, $k=0$ in the above results, the four-point function becomes the equal-weight stress-tensor multiplet correlator $\langle \mathcal{O}_{2}\mathcal{O}_{2}\mathcal{O}_{2}\mathcal{O}_{2}\rangle$. This correlator was first obtained in \cite{Arutyunov:2002ff} from the action of $\mathcal{N}=4$ gauged supergravity in seven dimensions on an $AdS_7$ background. Our calculation agrees with their result and the result obtained from the position space method \cite{longads7}.

\subsection{3D $\mathcal{N}=8$: 11D Supergravity on $AdS_4\times S^7$}\label{3dapplication}
\subsubsection{The Stress-Tensor Multiplet Four-Point Function}
In this section, we compute the holographic one-half BPS four-point function of operators with $k_i=2$ from eleven dimensional supergravity compactified on $AdS_4\times S^7$ at tree-level. The supergravity theory is conjectured to be dual to an $\mathcal{N}=8$ SCFT in three dimensions which describes the infrared limit of the effective theory on a large number $N$ of coincident $M2$-branes in flat space. An explicit realization of this effective theory is the ABJM theory \cite{Aharony:2008ug} with Chern-Simons level $k=1$ and  large $N$. The theory has  $OSp(4|8)$ superconformal symmetry\footnote{The full $\mathcal{N}=8$ superconformal symmetry of $k=1$ ABJM theory is not manifest at the classical level, but an enhancement from $\mathcal{N}=6$ to $\mathcal{N}=8$ is anticipated at the quantum level from string theory arguments.} which includes the conformal symmetry group $SO(3,2)$ and R-symmetry group $SO(8)$ as bosonic subgroups. To our knowledge, even this simplest four-point function has never been computed and written down in a closed form  in the literature. We cross it off from our to-do list by providing a solution to this correlator in Mellin space.

The one-half BPS operator $\mathcal{O}_{k=2}^{IJ}$ is the superconformal primary of the $\mathcal{N}=8$ stress-tensor multiplet. It has conformal dimension $\Delta=1$ and transforms as the symmetric-traceless representation $\mathbf{35}_c$ (the rank-two symmetric traceless product of the $\mathbf{8}_c$) under the $SO(8)$ R-symmetry group. In the bulk supergravity dual, it corresponds to a scalar field in $AdS_4$ with squared mass $m^2=-2$. From the supergravity selection rules, we find that only three fields can be exchanged in each channel, namely, the same scalar field itself, a vector field with dimension $\Delta=2$ in the representation $\mathbf{28}$ and a graviton field with dimension $\Delta=3$ in the singlet representation. However all the three fields have conformal twist $\tau=1$, the truncation condition of (\ref{trunccond}) therefore cannot be met and the Mellin amplitude will have an infinite series of poles. In the position space language, this means that the ``without really trying'' method of \cite{DHoker:1999aa} is no longer effective and the exchange diagrams can only be expressed as a infinite sum of $D$-functions. 

This property of the exchange diagrams presents some technical challenge even when we are working directly in Mellin space. Since the series of simple poles in the three channels do not truncate, we would need infinitely many parameters to parameterize the residues if we were to work with an ansatz such as used in the $AdS_5$ and $AdS_7$ cases. We postpone the analysis with such a general ansatz and take a more restrictive ansatz where the Mellin amplitudes of the exchange Witten diagrams are used. In total, the ansatz is a crossing-symmetric sum of the exchange Mellin amplitudes plus contact Mellin amplitudes. This reduces the variable coefficients to a finite set, with three of them tracking the contribution of the scalar, vector and graviton exchanges, and a few more for the contact diagrams. Such an ansatz is precisely what was used in the $AdS_5$ and $AdS_7$ position space method \cite{Rastelli:2017udc,longads7}. Note that the same ansatz for $AdS_4$ does not give us much mileage in position space  because of the difficulty to handle an infinite sum of $D$-functions. However, taking advantage of the simple structure of the Mellin amplitudes,  it is possible to obtain a closed form answer by solving the Mellin space version of the superconformal Ward identities. Let us now spell out the details.

As mentioned above, our ansatz for the full Mellin amplitude is a sum of the exchange diagram amplitudes and contact diagram amplitudes
\begin{equation}
\mathcal{M}(s,t;\sigma,\tau)=\mathcal{M}_{\text{s-exchange}}+\mathcal{M}_{\text{t-exchange}}+\mathcal{M}_{\text{u-exchange}}+\mathcal{M}_{\text{contact}}\;.
\end{equation}
The s-channel exchange amplitude $\mathcal{M}_{\text{s-exchange}}$ is comprised of the amplitudes from exchanging three fields
\begin{equation}
\mathcal{M}_{\text{s-exchange}}=\lambda_g\mathcal{M}_{\text{graviton}}(s,t)+\lambda_v(\sigma-\tau)\mathcal{M}_{\text{vector}}(s,t)+\lambda_s(4\sigma+4\tau-1)\mathcal{M}_{\text{scalar}}(s,t)
\end{equation}
where the factors $1$, $(\sigma-\tau)$, $(4\sigma+4\tau-1)$ are the R-symmetry polynomials associated with the irreducible representations of the exchanged fields. The numbers $\lambda_g$, $\lambda_v$ and $\lambda_s$ are the unknown coefficients that need to be fixed. Generally, the Mellin amplitude of an exchange Witten diagram contains a singular part which is a sum of simple poles and a regular part which is a polynomial. As was alluded to before, the singular part of the exchange Mellin amplitude has the same poles and residues as the Mellin space expression of an conformal block whose conformal dimension and spin are identical to those of the exchanged single-trace operator. On the other hand, since we have a contact part $\mathcal{M}_{\text{contact}}$ in the ansatz, we do not need to keep track of the polynomial piece in the exchange amplitudes. Such polynomial terms can just be swept into $\mathcal{M}_{\text{contact}}$ with a redefinition of the parameters. Therefore we only write down the singular pieces in $\mathcal{M}_{\text{graviton}}$, $\mathcal{M}_{\text{vector}}$, $\mathcal{M}_{\text{scalar}}$, and using the expressions in \cite{Fitzpatrick:2011hu} we have
\begin{eqnarray}
\nonumber\mathcal{M}_{\text{graviton}}&=&\sum_{n=0}^\infty\frac{3\sqrt{\pi}\cos[n\pi]\Gamma[-\frac{3}{2}-n]}{4n!\Gamma[\frac{1}{2}-n]^2}\frac{4 n^2-8 n s+8 n+4 s^2+8 s t-20 s+8 t^2-32 t+35}{s-(2n+1)}\;,\\
\nonumber \mathcal{M}_{\text{vector}}&=&\sum_{n=0}^\infty\frac{\sqrt{\pi}\cos[n\pi]}{(1+2n)\Gamma[\frac{1}{2}-n]\Gamma[1+n]}\frac{2t+s-4}{s-(2n+1)}\;,\\
\mathcal{M}_{\text{scalar}}&=&\sum_{n=0}^\infty\frac{\sqrt{\pi}\cos[n\pi]}{n!\Gamma[\frac{1}{2}-n]}\frac{1}{s-(2n+1)}\;.
\end{eqnarray}
In the above expressions we have appropriately symmetrized these amplitudes such that the residues in $\mathcal{M}_{\text{graviton}}$ and $\mathcal{M}_{\text{scalar}}$ are symmetric under exchanging $t\leftrightarrow u$ while the residues of $\mathcal{M}_{\text{vector}}$ are antisymmetric.  Here, $u=4-s-t$. These symmetry properties follow from the usual four-point amplitude kinematics. The t-channel and u-channel exchange amplitudes are related to the s-channel amplitude by crossing symmetry
\begin{eqnarray}
\mathcal{M}_{\text{t-exchange}}(s,t;\sigma,\tau)&=&\tau^2\mathcal{M}_{\text{s-exchange}}(t,s;\sigma/\tau,1/\tau)\;,\\
\mathcal{M}_{\text{u-exchange}}(s,t;\sigma,\tau)&=&\sigma^2\mathcal{M}_{\text{s-exchange}}(u,t;1/\sigma,\tau/\sigma)\;.
\end{eqnarray}
For the contact contribution $\mathcal{M}_{\text{contact}}$, we have the ansatz
\begin{equation}
\mathcal{M}_{\text{contact}}(s,t;\sigma,\tau)=\sum_{{\fontsize{4}{3}\selectfont\begin{split}{}&0\leq i,j\leq 2,\\{}&0\leq i+j\leq 2\end{split}}}\sum_{{\fontsize{4}{3}\selectfont\begin{split}{}&0\leq a,b\leq 1,\\{}&0\leq a+b\leq 1\end{split}}}\mu_{ij;ab}\;\sigma^i\tau^j s^a t^b\;,
\end{equation}
together with the crossing symmetry condition that
\begin{equation}\label{crossingid3d}
\mathcal{M}_{\text{contact}}(s,t;\sigma,\tau)=\tau^2\mathcal{M}_{\text{contact}}(t,s;\sigma/\tau,1/\tau)=\sigma^2\mathcal{M}_{\text{contact}}(u,t;1/\sigma,\tau/\sigma)\;.
\end{equation}
The fact that $\mathcal{M}_{\text{contact}}$ has degree two in $\sigma$ and $\tau$ is in agreement with the fact that $\mathcal{L}=2$, and its linear dependence on $s$ and $t$ is required by the flat space limit.

We now plug this ansatz into the Mellin space Ward identities. Solving these identities now becomes a bit nontrivial compared to the previous cases because we have infinitely many poles. Some observation can be made which allows us to solve all the coefficients in two steps. First notice all the poles of $s$, $t$ and $u$ at odd integer positions can only come from the exchange part. This is because the shift operations always shift the arguments by an even integer amount and the parity of poles are preserved. Requiring those poles to vanish gives
\begin{equation}
\lambda_v=-4 \lambda_s\;,\quad\quad \lambda_g=\frac{\lambda_s}{3}\;.
\end{equation}
These values are in agreement with the 3d $\mathcal{N}=8$ superconformal block \cite{Chester:2014fya}\footnote{To see this, let us take the residue of the leading pole at $s=1$ and perform the $t$-integral. The scalar contributes to the four-point function by $-2\pi^3\lambda_s U^{\frac{1}{2}}g^{\text{coll}}_{1,0}(V)$. Here $g_{\Delta,\ell}^{\text{coll}}(V)$ is the collinear block
\begin{equation}
g_{\Delta,\ell}^{\text{coll}}(V)\equiv g^{(0)}_{\Delta,\ell}(V)=(1-V)^{\ell}{}_2F_1\left(\frac{\Delta+\ell}{2},\frac{\Delta+\ell}{2},\Delta+\ell,1-V\right)\;.
\end{equation} Similarly, the vector contributes $\frac{\pi^3}{4}(\sigma-\tau)U^{\frac{1}{2}}\lambda_sg^{\text{coll}}_{2,1}(V)$ and the graviton contributes $-\frac{1}{256} \left(3 \pi ^3\right)U^{\frac{1}{2}}\lambda_s g^{\text{coll}}_{3,2}(V)$. Taking into account the normalization difference, $g^{\text{there}}=\frac{(\epsilon)_\ell}{4^\Delta(2\epsilon)_\ell}g^{\text{here}}$, we find the ratio $1:-1:\frac{1}{4}$ in (C.1-3) of \cite{Chester:2014fya}.}. To relate to the contact term parameters $\mu_{ij;ab}$, we need to look at the poles in the superconformal Ward identities\footnote{Note that we extract from all terms a common Gamma factor $\Gamma^2[2-s/2]\Gamma^2[2-t/2]\Gamma^2[2-u/2]$.} at $s+t=4,6,8$. Such poles come from the shifted Gamma functions in the reduced Mellin amplitudes. It is easiest if we first pick a specific $n$ for $\mathcal{M}_{\text{graviton}}$, $\mathcal{M}_{\text{vector}}$, $\mathcal{M}_{\text{scalar}}$ and obtain the residues. Then we resum in $n$ to get an expression, which needs to be canceled with the term coming from the contact part $\mathcal{M}_{\text{contact}}$.
Requiring that and the crossing identity (\ref{crossingid3d}), we obtain enough equations to solve all the $\mu_{ij;ab}$ coefficients with respect to $\lambda_s$. Plugging this solution into $\mathcal{M}_{\text{contact}}$, we get
\begin{equation}
\mathcal{M}_{\text{contact}}=\frac{\pi\lambda_s}{2}\left(-s-u\sigma^2-t\tau^2+4(t+u)\sigma\tau+4(s+u)\sigma+4(s+t)\tau\right)\;.
\end{equation}

Now let us fix the last coefficient $\lambda_s$. In \cite{Bastianelli:1999en} the three-point functions of the half-BPS operators were computed from supergravity 
\begin{equation}
\langle\mathcal{O}_2^{I_1}\mathcal{O}_2^{I_2}\mathcal{O}_2^{I_3} \rangle=\frac{2^{1/4}\sqrt{3\pi}}{N^{3/4}}\times \frac{\langle C^{I_1}C^{I_2}C^{I_3} \rangle}{x_{12}^2x_{13}^2x_{23}^2}\;
\end{equation}
where $\langle C^{I_1}C^{I_2}C^{I_3} \rangle$ is some R-symmetry tensor structure. In the small $U$ and small $V$ limit, the s-channel exchange of operator $\mathcal{O}_2$  contributes to the four-point function by (see Appendix B of \cite{longads7} for relating the $C$-symbols to the R-symmetry polynomials)
\begin{equation}
\frac{1}{2}\left(\frac{2^{1/4}\sqrt{3\pi}}{N^{3/4}}\right)^2(\sigma+\tau-\frac{1}{4})\; U^{1/2}g^{\text{coll}}_{1,0}(V)+\ldots\;.
\end{equation}
On the other hand, closing the contours in the inverse Mellin representation gives the following leading contribution in the $(\sigma+\tau-\frac{1}{4})$ R-symmetry channel
\begin{equation}
-2\pi^3\lambda_s\;(\sigma+\tau-\frac{1}{4})\times U^{1/2}g^{\text{coll}}_{1,0}(V)+\ldots\;.
\end{equation}
Matching these two expressions gives 
\begin{equation}
\lambda_s=-\frac{3\sqrt{2}}{4\pi^2N^{3/2}}\;.
\end{equation}

\subsubsection{Anomalous Dimension of $[\mathcal{O}_2\mathcal{O}_2]_{\Delta=2,\ell=0,\text{singlet}}$}\label{anom}
In \cite{Beem:2013qxa,Beem:2015aoa,Beem:2016wfs}, it was observed that  supergravity duals saturate the conformal bootstrap bounds for CFTs with maximal supersymmetry in four and six dimensions. A curious question therefore is if the same phenomenon persists also in three dimensions. In this subsection, we extract CFT data from the Mellin amplitude. We focus here on the anomalous dimension of the R-symmetry singlet scalar double-trace operators $[\mathcal{O}_2\mathcal{O}_2]_{\Delta=2,\ell=0,\text{singlet}}$. We compare the analytic result with the bound obtained from the numerical bootstrap. We find that the bootstrap estimation of the bound on the $1/C_T$ slope at large $C_T$ is reasonably close to the supergravity prediction.

Let us start with the leading disconnect piece of the four-point correlator
\begin{equation}
\mathcal{G}_{\text{disc}}=1+\sigma^2U+\frac{\tau^2U}{V}\;.
\end{equation}
Decomposing this correlator into different R-symmetry channels requires the use of the following $SO(8)$ R-symmetry polynomials which form a basis of degree-two polynomials
\begin{eqnarray}
\nonumber\mathbf{1}:&\quad\quad 1\;,\\
\nonumber\mathbf{28}:&\quad\quad \sigma-\tau\;,\\
\nonumber\mathbf{35}_c:&\quad\quad  \sigma + \tau -\frac{1}{4}\;,\\
\nonumber\mathbf{300}:&\quad\quad  \sigma ^2-2 \sigma  \tau  + \tau ^2-\frac{1}{3}\sigma-\frac{1}{3}\tau +\frac{1}{21}\;,\\
\nonumber\mathbf{567}_c:&\quad\quad  \sigma ^2 - \tau ^2-\frac{2}{5}\sigma+\frac{2}{5}\tau\;,\\
\mathbf{294}_c:&\quad\quad \sigma ^2+4 \sigma  \tau -\frac{2 \sigma }{3}+\tau ^2-\frac{2 \tau }{3}+\frac{1}{15}\;.
\end{eqnarray}
The projection of the disconnected correlator onto the R-symmetry singlet channel gives 
\begin{equation}
\mathcal{G}_{\text{disc, singlet}}=1+\frac{U(1+V)}{35V}\;.
\end{equation}
The first term of this singlet sector correlator is due to the exchange of the identity operator. The second term admits a decomposition into conformal blocks of double-trace operators $[\mathcal{O}_2\mathcal{O}_2]_{\Delta=2+2n+\ell,\ell,\text{singlet}}$ of the schematic form $:\mathcal{O}_2^{IJ}\square^n\partial^\ell\mathcal{O}_2^{IJ}:$ with $\ell$ even and their R-symmetry singlet superconformal descendants. For double-trace operators with $\Delta=2$, $\ell=0$, there is one {\it unique} such operator, namely, $[\mathcal{O}_2\mathcal{O}_2]_{\Delta=2,\ell=0,\text{singlet}}$. We can easily extract its zeroth order squared OPE coefficient from the above disconnected correlator and we get
\begin{eqnarray}
a^{(0)}_{n=0,\ell=0}&=&\frac{2}{35}\;.
\end{eqnarray}
In the small $U$ and small $V$ limit, the anomalous dimension $\gamma^{(1)}_{n=0,\ell=0}$ of this double-trace operator appears inside the term proportional to $U\log U$ of the four-point function
\begin{equation}
\begin{split}
\mathcal{G}_{\text{singlet}}(U,V)={}&A(V)U\log(U)+\ldots\;,\\
A(V)={}&\frac{1}{2}a^{(0)}_{n=0,\ell=0}\gamma^{(1)}_{n=0,\ell=0}g^{\text{coll}}_{2,0}(V)+\sum_{\ell\geq 2,{\text{ even}}}^\infty \sum_i \frac{1}{2}a^{(0)}_{n=0,\ell;i}\gamma^{(1)}_{n=0,\ell;i}g^{\text{coll}}_{2+\ell,\ell}(V)\;.
\end{split}
\end{equation}
The $U\log(U)$ term comes from the residue at the double pole $s=2$ in the inverse Mellin transformation, and the function $A(V)$ is further obtained by closing the contour of $t$. To extract the $\ell=0$ contribution, we use the following orthogonality property of ${}_2F_1$ \cite{Heemskerk:2009pn},
\begin{equation}
\begin{split}
{}&F_a(z)\equiv {}_2F_1(a,a;2a;z)\;,\\
{}&\oint_{z=0} \frac{dz}{2\pi i}z^{m-m'-1}F_{\Delta+m}(z)F_{1-\Delta-m'}(z)=\delta_{m,m'}\;.
\end{split}
\end{equation}
Evaluating the integrals, we find that the contact part of the Mellin amplitude contributes to $a^{(0)}_{n=0,\ell=0} \gamma^{(1)}_{n=0,\ell=0}$ by $-\frac{36\pi}{35}\lambda_s$ and the exchange part contributes to it by $\frac{106\pi}{35}\lambda_s$.
The total anomalous dimension therefore is
\begin{eqnarray}
\gamma^{(0)}_{n=0,\ell=0}=35\pi\lambda_s&=&-\frac{1120}{\pi^2}\frac{1}{C_T}\approx -113.5\frac{1}{C_T}\;
\end{eqnarray}
where $C_T=\frac{64\sqrt{2}}{3\pi}N^{3/2}$ in the convention of \cite{Chester:2014fya}. Supergravity hence yields the following large $C_T$ expansion for the conformal dimensions of the double-trace operators
\begin{eqnarray}
\Delta_0&\approx& 2-113.5\frac{1}{C_T}+\ldots\;.
\end{eqnarray}
In \cite{Chester:2014fya}, a numerical upper bound for the spin zero operator was reported, $\Delta_0^* \gtrsim 2.03-94.6/C_T+\ldots$. We see that this bound is compatible with the conjecture that supergravity should saturate the bootstrap bound, although the difference is still quite significant. This discrepancy can be partially explained by the slow convergence in the numerics for the scalar sector.\footnote{The same phenomenon was observed in \cite{Beem:2013qxa,Beem:2015aoa}.} Indeed, with more computational power a better estimation of the bound\footnote{We thank the authors of \cite{Agmon:2017xes} for providing their data before publishing.} is \cite{Agmon:2017xes}
\begin{equation}
\Delta_0^* \gtrsim 2.01-109/C_T+\ldots\;,
\end{equation}
and shows improved agreement with the supergravity result. Here for simplicity, we computed only the scalar singlet double-trace operator with the lowest conformal dimension. We took advantage of the absence of degeneracy of double-trace operators with these quantum numbers -- this is not true for double-trace operators with higher conformal dimensions, as we can see from the structure of the long multiplet (see, {\it e.g.}, Table 8 of \cite{Chester:2014fya}). To disentangle their contributions in the correlator, it is profitable to use superconformal blocks rather than just conformal blocks as used above -- we then only need to work with the superconformal primaries.

Let us end this subsection with a small comment. We recall that the same issue of double-trace operator degeneracy was encountered when analyzing the stress-tensor four-point function in 4d $\mathcal{N}=4$ SYM.\footnote{ An interesting fact to notice is that the t-channel exchange Witten diagram of a twist-two, spin-$J$ field gives the s-channel double-trace operators $:\mathcal{O}_{\Delta=2}\partial^\ell\mathcal{O}_{\Delta=2}:$  an anomalous dimension proportional to $\frac{1}{(\ell+1)(\ell+2)}$ for $\ell>J$ (see, {\it e.g.}, \cite{Alday:2017gde}). While all the exchanged single-trace fields have twist two, the correct answer for the anomalous dimension is proportional to $\frac{1}{(\ell+1)(\ell+6)}$ \cite{Dolan:2001tt}. The difference is accounted for precisely by the double-trace operator mixing.} In that problem, one can streamline the process of extracting anomalous dimensions by working with the function $\mathcal{H}$ that appears in the ``homogenous'' solution $\mathcal{K}$ as $\mathcal{K}\equiv R\mathcal{H}\equiv (1-\chi\alpha)(1-\chi\alpha')(1-\chi'\alpha)(1-\chi'\alpha')\mathcal{H}$. The key simplification exploited is the remarkable fact that multiplying the factor $R$ turns a bosonic conformal block $G_{\Delta,\ell}$ in $\mathcal{H}$ into a long superconformal block $G^{\text{long, super}}_{\Delta+4,\ell}$ in $\mathcal{G}$ \cite{Nirschl:2004pa}. Performing the conformal block decomposition on $\mathcal{H}$ then takes care of the superconformal descendants and automatically resolves the mixing problem in $\mathcal{G}$. A similar relation between bosonic conformal blocks and superconformal blocks also exists in 3d $\mathcal{N}=8$ \cite{Dolan:2004mu}. It would be interesting to explore a similar strategy for 3d $\mathcal{N}=8$, although the appearance of non-local differential operators in the solution to superconformal Ward identity makes it more non-trivial.

\section{Discussion}
In this paper we demonstrated how to impose superconformal constraints on four-point Mellin amplitudes. Our treatment works for all dimensions $3\leq d\leq 6$ where the position space superconformal Ward identity takes a universal form. This technique leads to a new way to efficiently compute holographic four-point functions and produces new results. For eleven dimensional supergravity compactified on $AdS_7\times S^4$ we computed the next-next-to-extremal four-point functions of operators with weights $k_1=n+k$, $k_2=n-k$, $k_3=k+2$, $k_4=k+2$. We showed that the four-point amplitude can be repackaged into a remarkably simple one-term auxiliary amplitude, in terms of a difference operator. We also computed the first one-half BPS four-point function for eleven dimensional supergravity compactified on $AdS_4\times S^7$ where the external operators are the superconformal primary of the 3d $\mathcal{N}=8$ stress-tensor multiplet. This four-point function is not accessible to the Mellin space approach employed in \cite{Rastelli:2016nze,Rastelli:2017udc,longads7} because the complicated  solution to the superconformal Ward identity in position space has no clear meaning in Mellin space. The position space method also falls short because the exchange Witten diagrams cannot truncate into finitely many contact Witten diagrams. Our new method hence provides a useful tool to initiate a systematic study of $AdS_4\times S^7$ holographic correlators. From the Mellin amplitude, we also extracted the anomalous dimension of the scalar R-symmetry singlet double-trace operator with the lowest conformal dimension.\footnote{Using our amplitude, a more systematic analysis was recently performed in \cite{Chester:2018lbz}, where the $1/C_T$ corrections to low-lying CFT data were obtained.} The result is compatible with the numerical bootstrap bound obtained in \cite{Chester:2014fya} and supports the conjecture that supergravity saturates the bootstrap bound at large central charge.  So far we have only restricted the application of our new technique to correlators with $\mathcal{L}=2$. It remains to be explored, especially for $AdS_4\times S^7$, when the four-point functions have more general weights and $\mathcal{L}>2$. The Mellin amplitudes of $AdS_4\times S^7$ four-point functions presumably should also admit some underlying structures that allow them to be captured by some simpler auxiliary amplitudes. It would be interesting to discover such structures in studying massive KK modes four-point functions. We have also omitted in this paper the study of the five dimensional superconformal CFT dual to IIA supergravity on the warped $AdS_6\times_w S^4 $. The case of 5d SCFT is different from the rest because the superconformal group is $F(4)$ and is not maximally supersymmetric\footnote{$F(4)$ can be viewed as the conformal version of 5d $\mathcal{N}=1$. 5d $\mathcal{N}=2$ would be maximally supersymmetric but it cannot be made conformal \cite{Nahm:1977tg}.}. We leave the treatment of Mellin amplitudes in non-maximally superconformal theories to a separate publication \cite{Zhou:2018ofp}.

\acknowledgments
X.Z. wants to thank Shai Chester, Wolfger Peelaers, Eric Perlmutter, Silviu Pufu, Leonardo Rastelli, Yifan Wang for useful conversations and is grateful to Wolfger Peelaers, Eric Perlmutter and Leonardo Rastelli for comments on the manuscript. The work of X.Z. is supported in part by NSF Grant PHY-1620628.

\bibliography{mellinward} 
\bibliographystyle{utphys}

\end{document}